\documentclass{aa}  
\usepackage{natbib}
\bibpunct{(}{)}{;}{a}{}{,} 

\usepackage{graphicx}
\usepackage{txfonts}

\usepackage{lipsum}
\usepackage{graphicx}
\usepackage{float}

\usepackage{adjustbox}
\usepackage{threeparttable}
\usepackage{booktabs, makecell}

\footnotesize \small 

\newcommand{\reduceme}{\mbox{R\raisebox{-0.35ex}{E}D%
\hspace{-0.05em}\raisebox{0.85ex}{uc}\hspace{-0.90em}%
\raisebox{-.35ex}{{m}}\hspace{0.05em}E}}

\usepackage[colorlinks = true,
            linkcolor = blue,
            urlcolor  = blue,
            citecolor = blue,
            anchorcolor = blue]{hyperref}

\usepackage{amstext}

\begin{document}

   \title{Near-IR narrow-band imaging with CIRCE at the Gran Telescopio Canarias: Searching for Ly$\alpha$-emitters at \mbox{$z \sim 9.3$}}

    \author{C.~Cabello
          \inst{1,2}\thanks{email: criscabe@ucm.es},
          J.~Gallego\inst{1,2},
          N.~Cardiel\inst{1,2},
          S.~Pascual\inst{1,2},
          R.~Guzmán\inst{1,2,3},
          A.~Herrero\inst{4,5},
          A.~Manrique\inst{6}, 
          A.~Marín-Franch\inst{7},
          J.M.~Mas-Hesse\inst{8},
          J.M.~Rodríguez-Espinosa\inst{4,5}
          \and
          E.~Salvador-Solé\inst{6}
          }

   \institute{
   Dept. de Física de la Tierra y Astrofísica, Fac. CC.Físicas, Universidad Complutense de Madrid, Plaza de las Ciencias 1, E-28040, Spain
   \and
   Instituto de Física de Particulas y del Cosmos (IPARCOS), Fac. CC. Físicas, Universidad Complutense de Madrid, Plaza de las Ciencias 1, E-28040 Madrid, Spain
   \and
   Dept. of Astronomy, University of Florida, Gainesville, USA
   \and
   Instituto de Astrofísica de Canarias, E-38205 La Laguna, Spain
   \and
   Dept. de Astrofísica, Universidad de La Laguna, E-38206 La Laguna, Spain
   \and
   Institut de Ciències del Cosmos. Universitat de Barcelona (UB-IEEC), Martí Franqués 1, E-08028 Barcelona, Spain.
   \and
   Centro de Estudios de Física del Cosmos de Aragón (CEFCA), Plaza san Juan 1,  E-44001 Teruel, Spain.
   \and
   Centro de Astrobiología - Dept. de Astrofísica (CSIC-INTA), E-28850 Madrid, Spain.}

   \date{Received date / 
   Accepted date }

  \abstract
   {Identifying very high-redshift galaxies is crucial for understanding the formation and evolution of galaxies. However, many questions still remain, and the uncertainty on the epoch of reionization is large. In this approach, some models allow a \mbox{double-reionization} scenario, although the number of confirmed detections at very high $z$ is still too low to serve as observational proof.
   }
   {The main goal of this project is studying whether we can search for Lyman-$\alpha$ emitters (LAEs) at \mbox{$z \sim 9$} using a narrow-band (NB) filter that was specifically designed by our team and was built for this experiment.
   }
   {We used the NB technique to select candidates by measuring the flux excess due to the Ly$\alpha$ emission. The observations were taken with an NB filter (full width at half minimum of~11~nm and central wavelength $\lambda_{c}$=~1.257~$\mu$m) and the CIRCE near-infrared camera for the Gran Telescopio Canarias.
   We describe a data reduction procedure that was especially optimized to minimize instrumental effects.
   With a total exposure time of 18.3~hours, the final NB image covers an area of $\sim$ 6.7~arcmin$^{2}$, which corresponds to a comoving volume of 1.1~$\times$~10$^{3}$~Mpc$^{3}$ at \mbox{$z = 9.3$}. 
   }
   {We pushed the source detection to its limit, which allows us to analyze an initial sample of 97 objects. We detail the different criteria we applied to select the candidates. The criteria included visual verifications in different photometric bands. None of the objects resembled a reliable LAE, however, and we found no robust candidate down to an emission-line flux of 2.9~$\times$~10$^{-16}$~erg~s$^{-1}~$cm$^{-2}$, which corresponds to a Ly$\alpha$ luminosity limit of 3~$\times$~10$^{44}$~erg~s$^{-1}$. We derive an upper limit on the Ly$\alpha$ luminosity function at \mbox{$z \sim 9$} that agrees well with previous constraints. 
   We conclude that deeper and wider surveys are needed to study the LAE population at the cosmic dawn.
   }
   {}


\authorrunning{C.~Cabello et al.}
\titlerunning{Near-IR narrow-band imaging with CIRCE at the GTC: Searching for Ly$\alpha$ emitters at \mbox{$z \sim 9.3$}}
\maketitle
%

\section{Introduction}\label{introduction}

Observing very high-$z$ galaxies is one of the main challenges in extragalactic astronomy because these early objects are extremely faint. Deriving their physical properties allows us to determine how and when they formed and in turn to better understand the mechanisms of formation and evolution of the galaxies that we observe today. In addition to this, finding \mbox{$z \sim 9$} objects is also essential to constrain the reionization epoch \citep{2005MNRAS.357.1348W,2007A&A...461..911C,2008flu..book..259E,2017ApJ...837L..21L,2021MNRAS.505.3336L}. 

The cosmic reionization is the transformation of the intergalactic medium (IGM) from neutral hydrogen at very high redshift into the ionized state we observe locally \citep{2019arXiv190706653W}. The transition follows a changing balance between the production rate of ionizing photons, with energies greater than 13.6~eV (wavelength, $\lambda<$~91.2~nm), and the recombination rate. The identification of the main sources and the way in which they ionized the IGM is still a matter of debate, but there are many sources of ionizing photons that could contribute to the reionization process, such as primitive stars (Population~III) and faint star-forming galaxies at very high $z$ \citep{2012ApJ...758...93F,2013ApJ...768...71R,2015ApJ...802L..19R}. Other alternative sources capable of producing reionizing photons include active galactic nuclei (AGN) or decaying elementary particles \citep{2010Natur.468...49R}. At the beginning, the sources were very scattered throughout the Universe, therefore reionization was not homogeneous, and initially, ionized bubbles began to form, which progressively grew and percolated. The Lyman-$\alpha$ (Ly$\alpha$) line emitted by very high-$z$ galaxies travels through the IGM and is absorbed by neutral hydrogen. If the amount of neutral hydrogen is high, this hampers the detection of the line. Thereby, the measurement of this line allows us to constrain the redshift range in which the IGM was fully ionized \citep{2016ARA&A..54..761S}.
The majority of the observed galaxies at low $z$ present low Ly$\alpha$ escape fractions. 
However, theoretically, luminous sources at very high $z$ can create large ionized bubbles that would allow the transmission of the Ly$\alpha$ photons \citep{2019ApJ...879...36F,2020ARA&A..58..617O}. These sources are optimum targets for follow-up spectroscopy, but luminous very high-$z$ galaxies are rare \citep{2020ApJ...904...50M}.
When the spectra of distant quasars are studied, the results suggest that at \mbox{$z \sim 6,$} the IGM is fairly ionized \citep{2015ApJ...811..140B}. 
Although there is some controversy, according to the observations we can establish the redshift range [6 -- 12] as the period in which the reionization process took place.
Owing to several contradictory results, the redshift at which the reionization began is unclear. The polarization measurements of the cosmic microwave background (CMB) from Planck Collaboration \citep{2020A&A...641A...6P} show a large optical depth due to Thompson scattering of electrons in the early Universe, suggesting a mid-point reionization redshift of \mbox{$z = 7.7 \pm 0.7$}. This is consistent with models in which reionization occurred relatively fast and late by an instantaneous process \citep{2015ApJ...802L..19R}.

However, there is strong evidence of substantial amount of ionizing radiation in the early Universe, as indicated by the detection of very high-$z$ galaxies at \mbox{$z \sim$ [8 -- 12]}. For instance, with $\sim$~4.3~h of exposure time using the Multi-Object Spectrometer For Infra-Red Exploration (MOSFIRE) spectrograph at the Keck Observatory,  \cite{2015ApJ...810L..12Z} confirmed the galaxy \mbox{EGSY-2008532660} (also known as EGSY8p7) as a Lyman-$\alpha$ emitter (LAE) at \mbox{$z = 8.68$} with a total Ly$\alpha$ flux of \mbox{1.0 $-$ 2.5~$\times$~10$^{-17}$~erg~s$^{-1}~$cm$^{-2}$}. On the other hand, the Hubble \mbox{WFC3/IR} slitless grism spectra of \mbox{GN-z11} analyzed by \cite{2016ApJ...819..129O} show that the Ly$\alpha$ break is redshifted to \mbox{$z = 11.10$}. This value points toward a strong Ly$\alpha$ emission just $\sim$~400~Myr after the Big Bang. Recently,  \cite{2018Natur.557..392H} have detected an emission line of doubly ionized oxygen from the galaxy MACS1149-JD at \mbox{$z = 9.11$} based on data from the Atacama Large Millimeter/submillimeter Array (ALMA). This star-forming galaxy was also studied with the X-shooter instrument, measuring a Ly$\alpha$ flux of  \mbox{4.3 $\pm$ 1.1~$\times$~10$^{-18}$~erg~s$^{-1}~$cm$^{-2}$}. All these studies improve our understanding of this key epoch of the Universe and support the idea of early galaxy formation. 
Of particular note are the spectroscopic data obtained with the Multi-Unit Spectroscopic Explorer instrument \citep[MUSE; ][]{2010SPIE.7735E..08B}, which shed light on the sources that are thought to have reionized the Universe \citep{2019A&A...628A...3D,2020A&A...638A..12K}. 
The unprecedented depth of the MUSE data allowed \cite{2018ApJ...865L...1M} to find 102~LAEs at \mbox{$2.9 $ < z < $6.7,$} reaching Ly$\alpha$ fluxes as faint as~$\sim$~8~$\times$~10$^{-19}$~erg~s$^{-1}~$cm$^{-2}$.

Measuring flux excesses in a narrow-band (NB) filter due to Ly$\alpha$ emission is one of the main techniques for detecting Ly$\alpha$ emitters at high redshift \citep{1995ApJ...449L...1P, 2016ARA&A..54..761S,2019SAAS...46..189O}. After lower-$z$ contaminants are rejected, accurate photometry in the NB image is needed to obtain Ly$\alpha$ luminosities and equivalent widths (EWs). The same technique has been employed by \cite{2019A&A...627A..84L} to search for LAEs at \mbox{$z = 8.8$}, using the infrared camera VIRCAM \citep{2015A&A...575A..25S} integrated at the 4.1~m survey telescope VISTA, and an NB filter (NB118) centered at 1.19~$\mu$m. With a total integration time of 168~h, they did not detect any galaxy candidate at \mbox{$z = 8.8$}.
When the survey is complete, the detection limit is expected to be \mbox{1.5 $-$ 1.7~$\times$~10$^{-17}$~erg~s$^{-1}~$cm$^{-2}$}, and the predicted probabilities of detecting these distant galaxies are roughly 10$\%$ for detecting one source and 1$\%$ for more than one source. 
This gives us an idea of how challenging the task of detecting galaxies on the verge of the epoch of reionization is, even though these observations are pushed to the limit.

Studying the evolution of the luminosity function (LF) of LAEs is another method to probe the Ly$\alpha$ transmission along the IGM at different redshifts. The decrease in UV LF with redshift suggests that high-$z$ galaxies produce fewer UV ionizing photons than lower-$z$ galaxies \citep{2010ApJ...723..869O, 2019ApJ...880...25B}.
Thus far, the Ly$\alpha$ LFs were only derived from observations at \mbox{$z \lesssim 7$} because only a few LAEs are detected at higher redshifts. 
Recently, \cite{2020ApJ...891L..10T} reported the most distant galaxy group composed of three galaxies at \mbox{$z \sim 7.7$}. The group was initially identified by NB imaging and was later spectroscopically confirmed with the MOSFIRE instrument. 
To compare with observations, \cite{2021ApJ...919..120M} modeled the evolution of the Ly$\alpha$ LF as a function of redshift, and average hydrogen neutral fraction. Their predictions show that at high $z$, when the Universe was more neutral, the characteristic Ly$\alpha$ luminosity decreases and the faint-end slope becomes steeper.

Because the observational data on the Universe at very \mbox{high $z$} are growing rapidly, numerous studies currently report accurate models and constraints for the reionization of the Universe \citep{2019ApJ...879...36F,2020ApJ...904..144J,2021MNRAS.502.5134B,2021MNRAS.508.1915P}. In this context, the ALBA team \citep{2017ApJ...834...49S} is actively following two parallel efforts: (1) the observational detection of very high-$z$ LAEs, and (2) the development of an Analytic Model of Intergalactic-medium and GAlaxy (AMIGA) \citep{2015ApJS..216...13M,2017ApJ...834...49S} evolution since the dark ages previous to reionization. Although Planck mission predicted one single reionization, the spectra of distant sources seem to indicate that reionization was extended and possibly nonmonotonous. 
AMIGA predictions are that the reionization very likely occurred in two stages separated by a short recombination period: a first stage at \mbox{$z \sim 10$} driven by the first generation of stars (i.e., Population~III stars), whose formation ended at this epoch as molecular cooling was quenched, and a second and definitive stage at \mbox{$z \sim 6$} that is due to young galaxies that formed at \mbox{$z > 6$}. The idea of double reionization was taken into account by several authors before \citep{2003ApJ...591...12C,2005ApJ...634....1F}, but AMIGA is the only model of galaxy formation developed so far that includes Population~III stars, normal galaxies, and AGNs, together with an inhomogeneous multiphase IGM. If confirmed, this scenario opens up the possibility that LAEs may be observed at \mbox{$z \sim 10$}. The comparison of number densities of LAEs and theoretical models is needed to confirm or reject the double-reionization scenario. 

The main goal of this project was to identify galaxies at the reionization epoch by deep near-infrared (\mbox{near-IR}) imaging with the CIRCE instrument at the Gran Telescopio Canarias (GTC). In particular, we search for LAEs at \mbox{$z \sim 9.3$} applying the NB technique in order to find observational proof to bring new light into the IGM and galaxy properties at the early Universe.
This paper is organized as follows. In Sect.~\ref{data}  we describe the instrumental setup of the observations and the ancillary data used in this study. We explain step by step the data reduction procedure in Sect.~\ref{datareduction}. The source extraction along with the depth and quality image characterization are described in Sect.~\ref{candidateidentification}. The selection and analysis of the candidates are performed in Sect.~\ref{finalsample}. We study the Ly$\alpha$ luminosity function in Sect.~\ref{LFsect}. Finally, in Sect.~\ref{summary} we present the summary and main conclusions of this paper. To complement this work, the analysis of the noise is described in Appendix~\ref{circe_errores}.

We use AB magnitudes throughout this paper. We assume a flat $\Lambda$CDM model with $\Omega_{\rm M}$=~0.30 and H$_{0}$~=~70 km~s$^{-1}$~Mpc$^{-1}$. 


\section{GTC near-IR and ancillary data}\label{data}

This work is based on deep \mbox{near-IR} imaging data obtained with the CIRCE instrument \citep{2018JAI.....750002E} on the GTC in La Palma, using an NB filter at 1.257~$\mu$m (hereafter referred to as NB1257). Because of the availability of public multiwavelength data, we have chosen a small region within the Extended Groth Strip (EGS) cosmological field. The EGS field is centered at $\alpha$~(J2000)~=~14h~17m~00s and $\delta$~(J2000)~=~+52$\degr~30\arcmin~00\arcsec$, corresponding to a high Galactic latitude ($b$~$\sim$~60$\degr$), therefore the Galactic foreground extinction is minimal. 
\begin{figure}[t]
  \resizebox{\hsize}{!}{\includegraphics{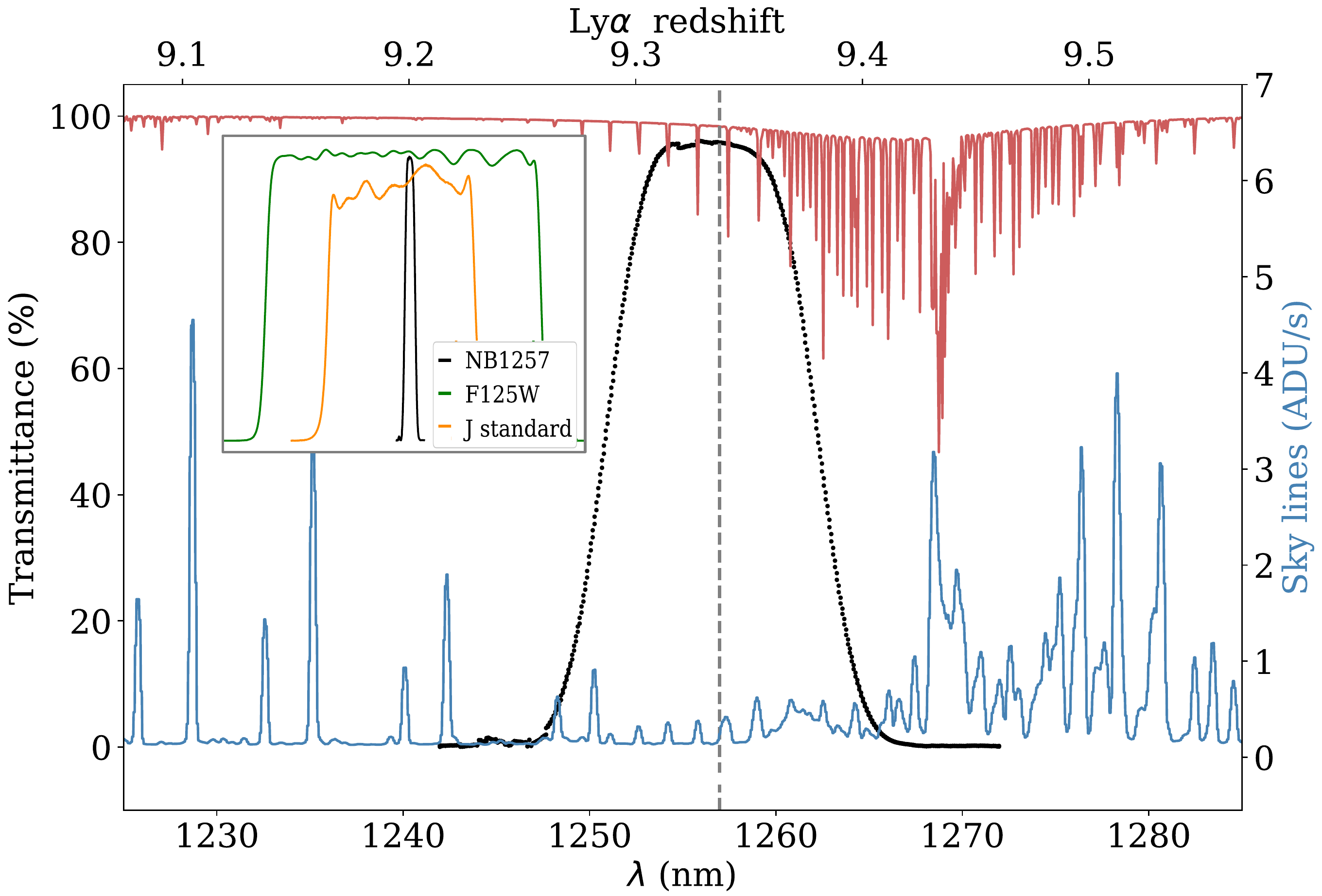}}
  \caption{Transmission curve of the NB1257 filter. The black dots show the empirical profile measured in the laboratory to guarantee FWHM~=~11~nm and central wavelength $\lambda_{c}$=~1.257~$\mu$m  (dashed vertical gray line) at cryogenic temperature. The sky emission of La Palma is overplotted in blue. This sky spectrum was obtained with the EMIR instrument at the GTC. The atmospheric absorption was computed based on the Cerro Paranal Advanced Sky Model and is plotted in red. The filter lies in a wavelength range with minimum sky continuum emission and OH emission of the atmosphere. The inset shows the comparison between the transmission curve of our NB1257 filter (solid black line) and the BB filters \mbox{WFC3/F125W} (solid green line) and J standard (solid orange line) from the Maunakea set of filters.}
  \label{transmission_band_filter}
\end{figure}
\begin{table}[t]\centering \caption[Detector features]{Characteristics of NB1257 filter }
\begin{threeparttable}
\begin{tabular}{l c}
\toprule\midrule
Parameter & Value\\
\midrule
FWHM & \phantom{00}11.0 $\pm$ 0.5 nm\\ 
Central wavelength ($\lambda_{c}$) & 1257.0 $\pm$ 2.0 nm \\ 
Diameter & \phantom{00}68.0 $\pm$ 0.1 mm  \\ 
Thickness & \phantom{000}6.0 $\pm$  0.1 mm \\ 
\bottomrule\addlinespace[1ex]
\end{tabular}
\end{threeparttable}
\label{table_filter}
\end{table}
The All-wavelength Extended Groth strip International Survey (AEGIS; \citealp[]{2007ApJ...660L...1D}) studied this region of the sky by employing four orbiting telescopes and four ground-based telescopes, including Keck spectroscopy \citep{2013ApJS..208....5N}. 
In addition, both the CANDELS \citep{2011ApJS..197...35G, 2011ApJS..197...36K} and \mbox{3D-HST} \citep{2011ApJ...743L..15V,2012ApJS..200...13B,2014ApJS..214...24S,2016ApJS..225...27M} surveys provide extensive satellite data (in particular from \mbox{HST/WFC3}) and derived subproducts such as photometric redshifts and stellar masses. 


\subsection{ALBA NB1257 filter at the CIRCE camera}

The Canarias InfraRed Camera Experiment (CIRCE) is a \mbox{near-IR} camera operating in the [1 -- 2.5] micron wavelength range, designed to be used as a visitor instrument at GTC. The instrument is equipped with a 2048~$\times$~2048~pixels engineering-grade HAWAII-2RG detector with gain = 5.3 e$^{-}$ /ADU and readout noise $\sim$~45~e$^{-}$ RMS. The plate scale is 0.1~$\arcsec$/pixel and the total field of view (FOV) is 3.4~$\times$~3.4~arcmin$^{2}$. The detector is subdivided into 32 independent channels for quick readout, although 2 channels are inactive.

The NB1257 filter was specifically designed by the ALBA team for this project. During several months, the ALBA team and SCHOTT technicians worked together to optimize the manufacture and the overall performance of the filter. It has a central wavelength of 1257~nm (corresponding to a redshifted  Ly$\alpha$ emission from sources at \mbox{$z = 9.34$}), and a full width at half maximum (FWHM) of $\Delta\lambda = 11$~nm. The filter width samples Ly$\alpha$ emission in the redshift range \mbox{$z =$ [9.29 -- 9.38],} and it was intended to match a region of low OH emission from the night sky. In addition to this, it presents a transmission of nearly $\sim~99\% $.

Table \ref{table_filter} shows the main features, but detailed information about the filter is provided by \cite{2018SPIE10702E..23B}. To prove the spectral uniformity of the bandpass filter, the measurements were taken at five different points (center point plus four points about 5--10~mm from the edge). The uniformity of the central wavelength was 0.14\% and the FWHM showed a variation of 0.52\%.  The maximum transmission of the filter was also uniform because the variation over the five points was only 0.33\%, and with values always higher than 98\% \citep{2019SPIE11180E..87H}.

\begin{table*}[ht]\centering \caption[OB summary]{Summary of the observing blocks.}
\begin{threeparttable}
\begin{tabular}{l c c c c c c c}
\toprule\midrule
\begin{tabular}[c]{@{}c@{}} OB \\ \\\phantom{++}(1)\end{tabular} &
\begin{tabular}[c]{@{}c@{}}Dithering pattern\\(ramps $\times$ t $\times$ frames) \\(2)\end{tabular}&
\begin{tabular}[c]{@{}c@{}}Date\\ (yyyy-mm-dd) \\(3)\end{tabular}&
\begin{tabular}[c]{@{}c@{}}Airmass\\  \\ (4)\end{tabular} &
\begin{tabular}[c]{@{}c@{}}Total exp. \\ (s)\\ (5)\end{tabular}& 
\begin{tabular}[c]{@{}c@{}}Seeing \\ ($\arcsec$)\\ (6)\end{tabular}& 
\begin{tabular}[c]{@{}c@{}} log(RobustStd) \\ \\ (7)\end{tabular}& 
\begin{tabular}[c]{@{}c@{}}Quality\\  \\ (8)\end{tabular}\\ \hline
OB0001 & 3  $\times$ 100  $\times$ 11  & 2016-05-20    & 1.10 & 3300 & 0.72  & 0.63 & $\checkmark$\\ 
OB0002 & 3  $\times$ 100  $\times$ 11  & 2016-05-20    & 1.11 & 3300 & 0.70  & 0.70 & $\checkmark$\\ 
OB0003 & 3  $\times$ 100  $\times$ 11  & 2016-05-21    & 1.16 & 3300 & 0.66  & 0.70 & $\checkmark$\\ 
OB0004 & 5  $\times$  60  $\times$ 11  & 2016-06-23    & 1.11 & 3300 & 0.65  & 0.60 & $\checkmark$\\ 
OB0005 & 5  $\times$  60  $\times$ 12  & 2016-06-23    & 1.17 & 3600 & 0.68  & 0.59 & $\checkmark$\\ 
OB0006 & 5  $\times$  60  $\times$ 11  & 2016-06-25    & 1.11 & 3300 & 0.54  & 0.51 & $\checkmark$\\ 
OB0007 & 5  $\times$  60  $\times$ 11  & 2016-06-25    & 1.18 & 3300 & 0.87  & 0.49 & $\checkmark$\\ 
OB0008 & 5  $\times$  60  $\times$ 11  & 2016-07-20    & 1.30 & 3300 & 0.71  & 0.63 & $\checkmark$\\ 
OB0009 & 5  $\times$  60  $\times$ 11  & 2016-07-21    & 1.39 & 3300 & 0.61  & 0.62 & $\checkmark$\\ 
OB0010 & 5  $\times$  60  $\times$ 11  & 2016-07-22    & 1.27 & 3300 & 0.70  & 0.53 & $\checkmark$\\ 
OB0011 & 5  $\times$  60  $\times$ 11  & 2017-04-07    & 1.14 & 3300 & 0.66  & 0.44 & $\checkmark$\\ 
OB0012 & 5  $\times$  60  $\times$ 11  & 2017-04-07    & 1.10 & 3300 & 0.57  & 0.43 & $\checkmark$\\ 
OB0013 & 5  $\times$  60  $\times$ 11  & 2017-04-07    & 1.11 & 3300 & 0.75  & 0.37 & $\checkmark$\\  
OB0014 & 5  $\times$  60  $\times$ 11  & 2017-04-07    & 1.17 & 3300 & 0.69  & 0.39 & $\checkmark$\\ 
OB0015 & 5  $\times$  60  $\times$ 11  & 2017-06-03    & 1.30 & 3300 & 1.04  & 0.70 & $\checkmark$\\ 
OB0016 & 5  $\times$  60  $\times$ 11  & 2017-06-03    & 1.11 & 3300 & 0.71  & 0.95 & --\\ 
OB0017 & 5  $\times$  60  $\times$ 11  & 2017-06-03    & 1.10 & 3300 & 0.70  & 0.91 & --\\ 
OB0018 & 5  $\times$  60  $\times$ 11  & 2017-06-03/04 & 1.13 & 3300 & 0.90  & 0.79 & --\\ 
OB0019 & 5  $\times$  60  $\times$ 11  & 2017-06-10    & 1.39 & 3300 & 0.84  & 0.74 & $\checkmark$\\ 
OB0020 & 5  $\times$  60  $\times$ 11  & 2017-06-10    & 1.67 & 3300 & 0.84  & 0.66 & $\checkmark$\\ 
OB0021 & 5  $\times$  60  $\times$ 12  & 2017-06-13    & 1.33 & 3600 & 0.83  & 0.60 & $\checkmark$\\ 
OB0022 & 5  $\times$  60  $\times$ 11  & 2017-06-13    & 1.57 & 3300 & 0.83  & 0.59 & $\checkmark$\\ 
OB0023 & 5  $\times$  60  $\times$ 09  & 2017-07-05    & 1.55 & 2700 & 1.13  & 0.62 & $\checkmark$\\ 
\bottomrule\addlinespace[1ex]
\end{tabular}
\tablefoot{Columns: (1):~Observing block IDs. (2):~Dithering pattern; number of ramps (the number of times to repeat the readout scheme selected), exposure time in each ramp in seconds, and number of dithered positions. (3):~Observation date (yyyy-mm-dd). (4):~Mean airmass between initial and final exposures. (5):~Total exposure time in seconds. (6):~Seeing in arcseconds. (7):~Logarithm of the robust standard deviation of the central part of the image (see more details in Appendix \ref{circe_errores}). (8):~Quality of the OBs: checkmarks indicate OBs with the expected quality that were finally used in this work.}
\end{threeparttable}
\label{tableOB}
\end{table*}

Figure~\ref{transmission_band_filter} shows the transmission curve of the NB1257 filter (black dots). We specifically targeted at \mbox{$z = 9.34$} (dashed vertical gray line) as the best compromise that would optimize the detection of LAEs considering our model predictions for the first reionization epoch (around \mbox{$z \sim 10$}) while minimizing the negative effects of the sky background (blue line) and atmosphere absorptions (red line). The telluric transmission spectrum was computed with \textsc{SkyCalc}, a web application based on the Cerro Paranal Advanced Sky Model\footnote{\url{https://www.eso.org/observing/etc/doc/skycalc/helpskycalc.html}}. In order to have a reliable measure of the sky emission in the \mbox{near-IR} range, we extracted the sky spectrum from commissioning data obtained with the EMIR instrument at the GTC \citep{2016ASPC..507..297G}, applying the \textsc{pyemir}\footnote{\url{https://pyemir.readthedocs.io/en/latest/}} reduction package \citep{2019ASPC..523..317C}. In particular, the sky spectrum was obtained in the J band on the night of 30 May 2018, and a slit width of 0.8$\arcsec$ was used during the observation. The redshift corresponding to the Ly$\alpha$ emission is in the top X-axis of Fig.~\ref{table_filter}.
Our NB1257 filter covers an atmospheric window with excellent conditions in terms of sky transmission. Although there are other atmospheric windows with similar sky properties (e.g., the window corresponding to Ly$\alpha$ emission at redshifts \mbox{$z =$ [8.7 -- 8.8]}), we selected the best window at the highest possible redshift, aiming to provide observational proof of the double-reionization scenario predicted by the AMIGA model.
Because the Ly$\alpha$ emission of sources at \mbox{$z =$ 10} falls in a wavelength range dominated by atmospheric absorptions, we targeted the Ly$\alpha$ line of \mbox{$z = 9.34$} sources. 
The observations reveal that the number of LAEs observed at \mbox{$z > 6.5$} decreases rapidly. This is explained by the increase in neutral hydrogen in the IGM \citep{2012ApJ...752..114S,2014ApJ...795...20S}, but the recent discovery of LAEs at \mbox{$z \sim 9$} seems to contradict this reionization scenario. 
Nonetheless, the double reionization can explain these detections because it predicts a high transparency of the sky for the Ly$\alpha$ photons due to a first peak of hydrogen ionization at \mbox{$z =$ [9 -- 10]}.
On the other hand, if the detection of LAEs at these very high redshifts is confirmed, the cosmological consequences of the finding would be of extraordinary relevance. This means that these observations should be preferred over other apparently more conservative observations at lower redshifts.

The instrumental setup during the observations was composed of two filters: the NB1257 filter, and a broad-band (BB) filter in the J-band region (see Fig.~\ref{transmission_band_filter} for the respective filter transmissions) as a blocking measure to avoid the light that comes from beyond our intended range. The CIRCE J-band filter is a standard filter from the Maunakea set \citep{2002PASP..114..180T}. Because our target galaxies are faint, we expected to detect only the emission line without reaching the continuum level in most cases. The transmission curve of the filter has an imperfect box-shaped profile, therefore the final observed emission-line flux will depend on the exact location of the emission line within the wavelength range that is covered by the filter \citep{2008ApJ...677..169V}. In particular, when a bright emission line lies close to the wavelength cutoff, the combination of low transmittance and strong flux is indistinguishable from a fainter emission line that is well centered in the filter.


 \begin{figure}[ht]
  \resizebox{\hsize}{!}{\includegraphics{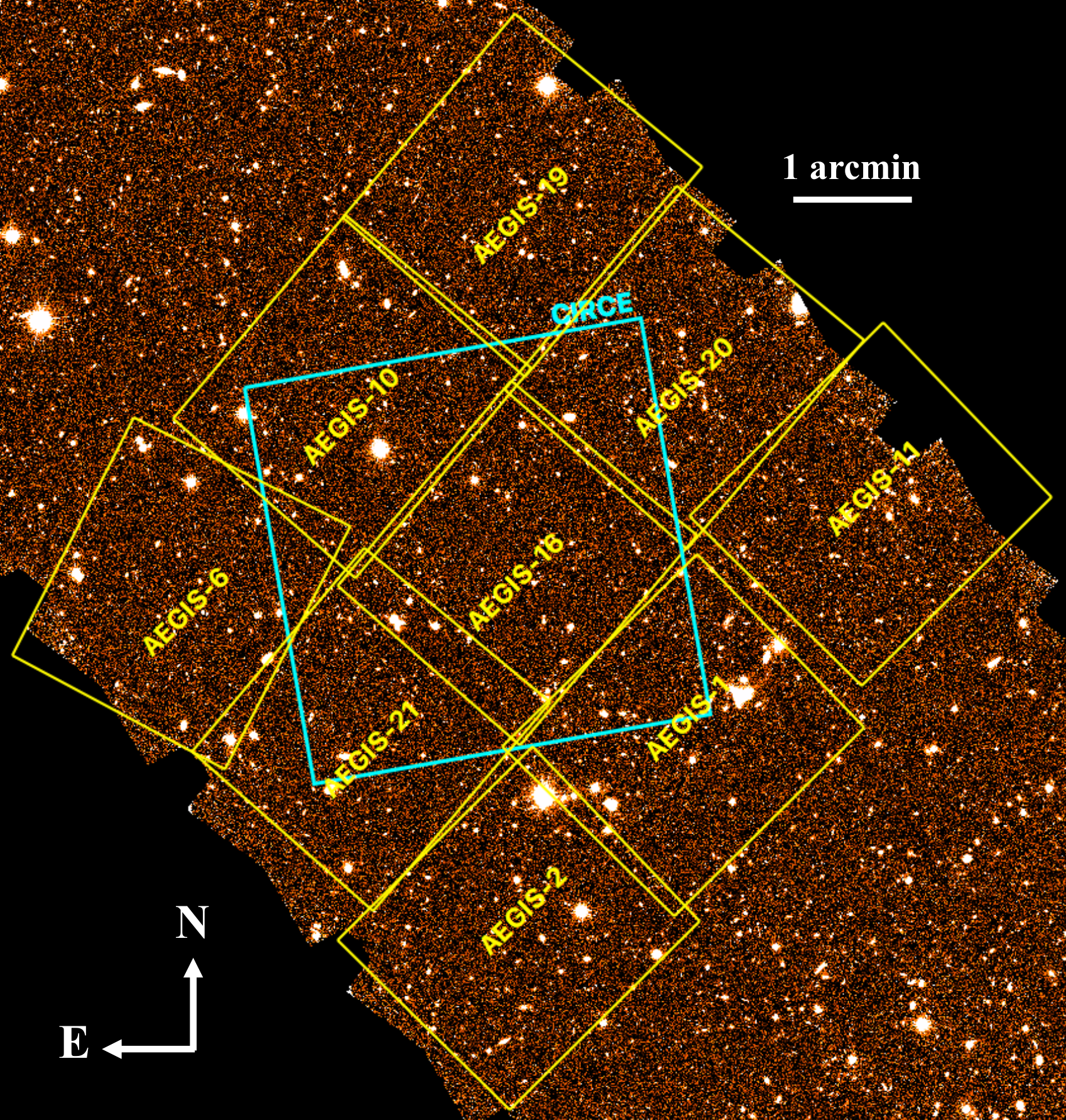}}
  \caption{Observation region within the EGS where the AEGIS tiles covered by \mbox{HST/WFC3} are shown with yellow boxes. The FOV of our CIRCE pointing is marked with a blue box of 3.4$\arcmin$~$\times$~3.4$\arcmin$. Image background: \mbox{WFC3/F125W} from the CANDELS survey. Position: $\alpha$~(J2000)~=~14h~19m~44.67s, $\delta$~(J2000)~=~$+52\degr~53\arcmin~11.5\arcsec$. Orientation: North is up, and east is to the left.}
  \label{fov_aegis}
\end{figure}

\subsection{CIRCE NB imaging}

The final pointing of our CIRCE observations was selected after a visual inspection of the individual \mbox{WFC3/F140W} tiles from the \mbox{3D-HST} survey in the EGS field. AEGIS-16 is one of the tiles that is covered by the HST, and our CIRCE pointing is centered on it. 
Based on the feasibility of the observations, the goal was to select a sky region without bright sources, except for a single star that was used for alignment purposes. The bright star in the field (magnitude in the \mbox{g band} of \mbox{m$_{\rm g}$ $=$ 14.9 mag}) was essential to carry out the data reduction (see Section~\ref{datareduction}).
The field of view of our CIRCE pointing (\mbox{$3.4\times3.4$}~arcmin$^{2}$, centered at $\alpha$~(J2000)~=~14h~19m~44.67s, $\delta$~(J2000)~=~$+52\degr~53\arcmin~11.5\arcsec$) is represented as a blue box in Fig.~\ref{fov_aegis}, and the surrounding AEGIS tiles observed by \mbox{HST/WFC3} are marked with yellow boxes as a reference. The background image displayed in this figure corresponds to the \mbox{WFC3/F125W} filter (CANDELS survey), which encompasses the wavelength coverage of our NB1257 filter.

The CIRCE data were obtained between May 2016 and July 2017. They consist of 23 pointings, and because the observations for this program were carried out in service mode, they were conveniently arranged in observing blocks (OBs) of $\sim$~1 hour of exposure time. Details of the OBs are given in Table~\ref{tableOB}. A standard dithering technique (shift + coadd) was used to deal with the cosmetic defects of the detector. Thus, by combining multiple images of a target at slightly different positions on the detector, one can compensate for detector artifacts (such as dead pixels). Each OB consisted of 11 or 12 dithered locations around the central coordinates of the target field, with offsets of a few dozen arcseconds ($\sim$~1/10 of the CIRCE FOV). At each particular telescope pointing within the dithering pattern, several consecutive exposures (three or five) were obtained before moving to the next location in that pattern. In turn, each exposure was performed in correlated double sampling (CDS) mode by reading the detector twice, once at the beginning and once at the end of the predefined exposure time. The subtraction of these two readings in CDS mode leads to the scientific image that is commonly referred to as the ramp.
The raw FITS images provided by the telescope contain the individual readings carried out at each telescope location within the dithering pattern (where each reading is stored as an independent FITS extension). As an example, for the first OB (see Table~\ref{tableOB}), the completion of this observing block provided 11 raw FITS files containing six extensions, that is, three exposures of 100 seconds each.
We did not need specific calibration images because no large objects were present in the images. In this case, the science images themselves were used for flat fielding and sky subtraction.


\subsection{HST/WFC3 near-IR imaging and ancillary data}

The \mbox{3D-HST} survey provides very deep imaging in a large number of photometric filters. In particular, it includes \mbox{HST/WFC3} data taken with the F125W, F140W, and F160W filters for~$\sim$~41200 objects over an area of~$\sim$~206~arcmin$^{2}$ within the EGS field. Because very high-$z$ galaxies at \mbox{$z \sim 9.3$} will be dropouts in all photometric bands bluer than J, we used the F125W data as our reference BB image. This comparison should allow us to select targets with a flux excess in the NB1257 filter. The transmission curve of the F125W filter is plotted in green in the inset of Fig.~\ref{transmission_band_filter}, where the narrowness of our NB1257 filter is highlighted over the other filters. The great advantage of the \mbox{3D-HST} treasury program is the spatially resolved low-resolution spectroscopic data taken by the \mbox{WFC3/G141} grism (wavelength coverage of~1.1--1.7~$\mu$m). 

In order to exploit different catalogs that are publicly available, we also gathered the archival data from the Cosmic Assembly Near-infrared Deep Extragalactic Legacy Survey (CANDELS) for the EGS field, which provides robust photometric redshifts and stellar masses inferred from spectral energy distributions \citep{2017ApJS..229...32S}.  


%
\begin{figure*}[ht]
  \resizebox{\hsize}{!}{\includegraphics{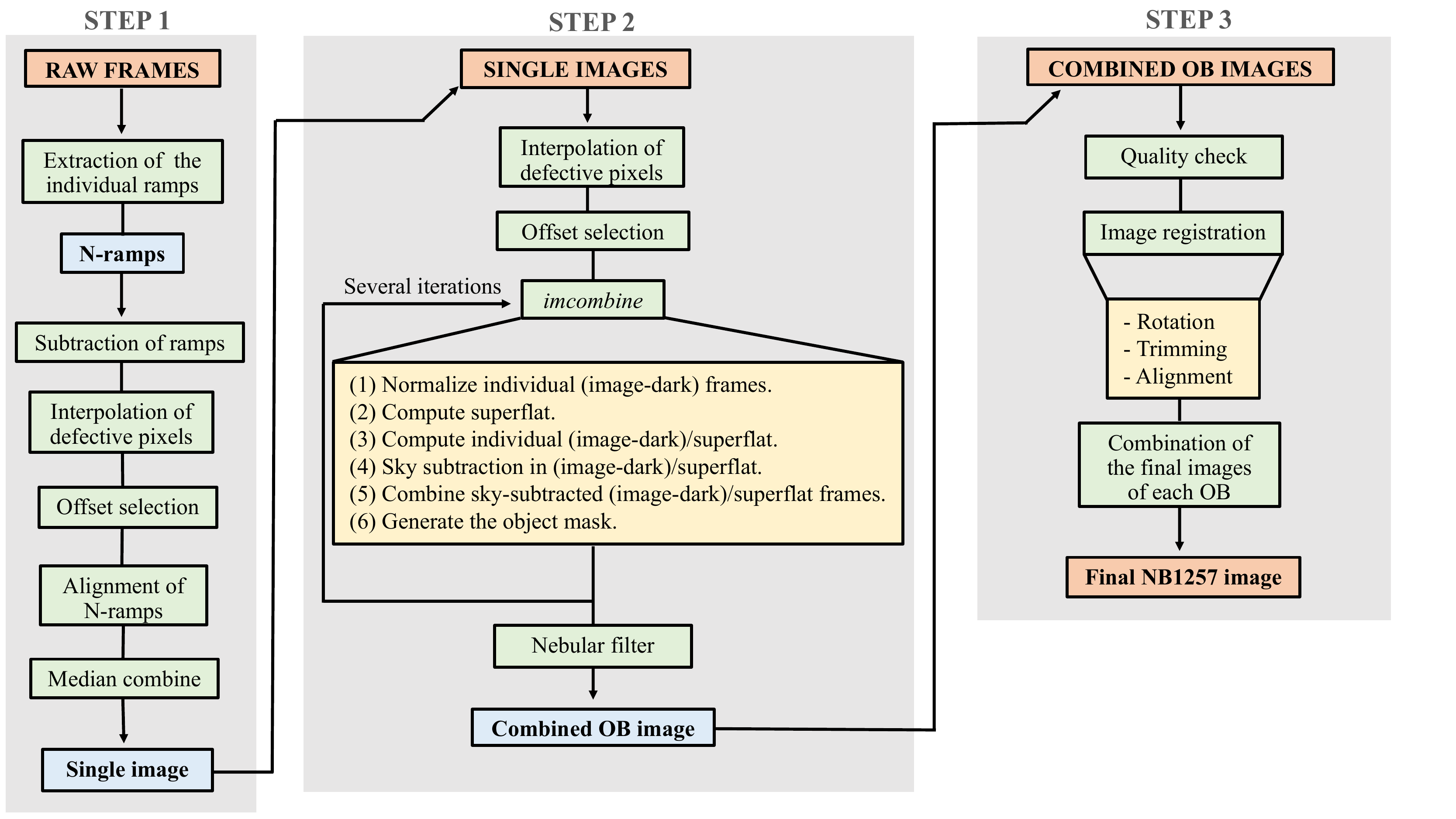}}
  \caption{Flowchart of the CIRCE data reduction procedure. Because the procedure is complex, we divided it into three steps as explained in detail in the text. Starting from the raw frames provided by the telescope and generating the single images (step 1), and after performing some additional manipulations, we obtain the combined images of each OB (step 2), and finally, the deep NB1257 image (step 3).}
  \label{DR_diagram}
\end{figure*}

\section{Data reduction}\label{datareduction}

We detail in this section the data reduction procedure that we used to generate the final deep NB image. We used different tools and routines to reduce the NB data, such as Python codes developed by ourselves (code available in GitHub\footnote{\url{https://github.com/criscabe/CIRCE}}), some tasks of IRAF (Image Reduction and Analysis Facility\footnote{\url{http://iraf.noao.edu/}}, \citealp{1986SPIE..627..733T,1993ASPC...52..173T}),  the {\sc cleanest\footnote{This software utility was initially developed by \cite{cardiel} as part of the \reduceme\ package, and it is at present available as a stand-alone program. See \url{https://cleanest.readthedocs.io/en/latest/}}} software to interpolate defective pixels, and  the {\sc imcombine}\footnote{\url{https://github.com/nicocardiel/imcombine}} software, a Fortran program to reduce \mbox{near-IR} images. All of this allows us to control the reduction process step by step, as shown in the flowchart in Fig.~\ref{DR_diagram}, where the whole data reduction process was subdivided into three main steps:
\begin{itemize}
    \item[$\bullet$] Step\,1: Reduction of each raw image by extracting and combining the different ramps, and generating a single image corresponding to each particular location within the dithering pattern of the considered OB.
    \item[$\bullet$] Step\,2: Combination of the resulting images from \mbox{step\,1} corresponding to the same OB. This process generates a single combined OB image.
    \item[$\bullet$] Step\,3: Combination of the resulting images from \mbox{step 2} corresponding to the valid OBs. 
\end{itemize}

Each of these steps is described in more detail in the following subsections.


\subsection{Step 1: Single image at each telescope pointing}\label{linearization}

\begin{figure}[t]
  \resizebox{\hsize}{!}{\includegraphics{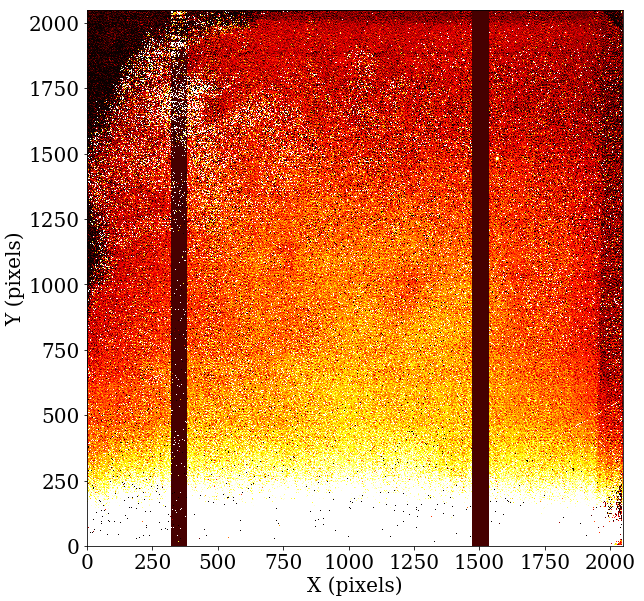}}
  \caption{Ramp image taken with the CIRCE detector. Exposure time: 100~s. Size: 2048~$\times$~2048~pixels. The borders of the detector are not useful because of the poor cosmetic. There are many defective pixels and two nonfunctional channels (vertical dark columns located at X coordinates $\sim$~350 and 1500).}
  \label{ramp}
\end{figure}

As previously explained, the subtraction of consecutive detector readings leads to ramp images of the predefined exposure time. An example of one of these ramps is shown in Fig.~\ref{ramp}. \mbox{Near-IR} detectors typically exhibit more cosmetic defects than their optical CCD counterparts, and, as is evident from this figure, the CIRCE detector was not an exception. Two of the 32 detector channels were not useful (vertical columns from 321 to 384 and from 1473 to 1536), while the detector had a non-negligible number of bad pixels, particularly near the borders and in the upper left corner.

Although we started the observing program requesting OBs with individual integrations of 100 seconds per ramp (OB0001 to OB0003) and three ramps per telescope pointing (i.e., location within the dithering pattern), we discovered that this strategy was not adequate for our purposes after examining in detail the straightforward combination of the three ramps in each raw frame. Because the GTC did not provide guiding correction for CIRCE exposures, the point spread function\textit{} (PSF) of the bright star visible in the field of view sometimes appeared slightly elongated after 300 seconds, which meant that the telescope tracking was not always good enough to prevent the drift of the telescope during the integration at each telescope location. To alleviate this problem, two actions were taken: 1)~we modified the observation strategy for the remaining OBs of this program by reducing the exposure time per ramp from 100 to 60 seconds and by subsequently increasing the number of ramps (exposures at each telescope location) from three to five in each raw frame (to preserve a total exposure time of 300 seconds at each position within dithering pattern), and 2)~we split the individual ramps within each raw frame to measure and apply any potential offset between ramps before their combination. The second action led to the introduction of this step~1 in the reduction procedure.

As graphically described in Fig.~\ref{DR_diagram}  (left panel), this step took an individual raw frame as input and generated a single image resulting from the refined combination of the available ramps (three or five). Initially, the individual ramps were extracted separately. Because it was necessary to remove the background level to clearly see even the brightest object in each image, we used as background level for each ramp the corresponding ramp in the next raw frame within the dithering pattern (ramps within the same raw frame cannot be subtracted for this purpose because all of them corresponded to the same telescope pointing).
   
Because the goal of this reduction step was to detect any potential offset between the different ramps due to the imperfect telescope tracking, we chose the bright star present in the field to apply a 2D cross-correlation technique. Considering the relatively high number of defective pixels in the CIRCE detector, we found it essential to interpolate these defective pixels in order to avoid biased results in the cross-correlation procedure. This cosmetic process was carried out manually using the specialized software {\sc cleanest} to remove bad pixels in an image subregion around the location of the bright star. The measured offsets derived by applying the 2D cross-correlation technique were rounded to integer values. Although these values were rounded to zero for many cases, the offsets amounted up to three pixels in a particular direction for some ramps. In these cases, the offsets were corrected to obtain perfectly aligned ramps. To have an idea of the uncertainties associated with the measured offsets, we repeated this process several times by situating the bright star at different positions within the window of 41~$\times$~41~pixels that we employed when we computed the cross-correlation function. In all cases, the measured errors were always smaller than 1~pixel (i.e., 0.1~arcsec).

After all the ramps within each raw frame were properly aligned, we combined them using the median value for each pixel, which allowed us to reject some defective pixels. However, the offsets were zero in many cases, and therefore this median combination did not remove the majority of the bad pixels.

\subsection{Step 2: Combined OB image} \label{imcombine_section}

By repeating the previously described step 1 for all the individual raw images (11 or 12, depending on the considered OB) of each observing block, we ended up with the same number of single images, one at each location within the dithering pattern. The main objective of step 2 here described (see Fig.~\ref{DR_diagram}, middle panel) was to align and combine these individual images to generate a single combined image for each OB. The bulk of this process was carried out using {\sc imcombine}, as described below. One of the inputs of this program is the offsets between individual images, but the coordinates of the header were not accurate enough to use them. Therefore, the relative offsets between the single images were obtained by following the same procedure as for computing the offsets between ramps, that is, by using the 2D cross-correlation technique in manually cleaned image subregions of $\sim$~100~$\times$~100~pixels around the two brightest stars (in this case, each single image was now a combination of three or five ramps, which allowed the application of the cross-correlation technique on the second brightest star as well). For each OB, the offsets were always measured relative to the first single image in the sequence, which provided the actual dithering pattern followed by the telescope when carrying out the observations of that particular OB.
Although the OB0023 was initially composed of 11 frames because the faint
star was barely visible in two single images and the bright star fell in one
of the inactive channels of the detector,  we discarded these two single images (see Table~\ref{tableOB}, Col.~2) because it is not possible measuring the corresponding offsets.

\begin{figure}[t]
\centering
  \resizebox{\hsize}{!}{\includegraphics{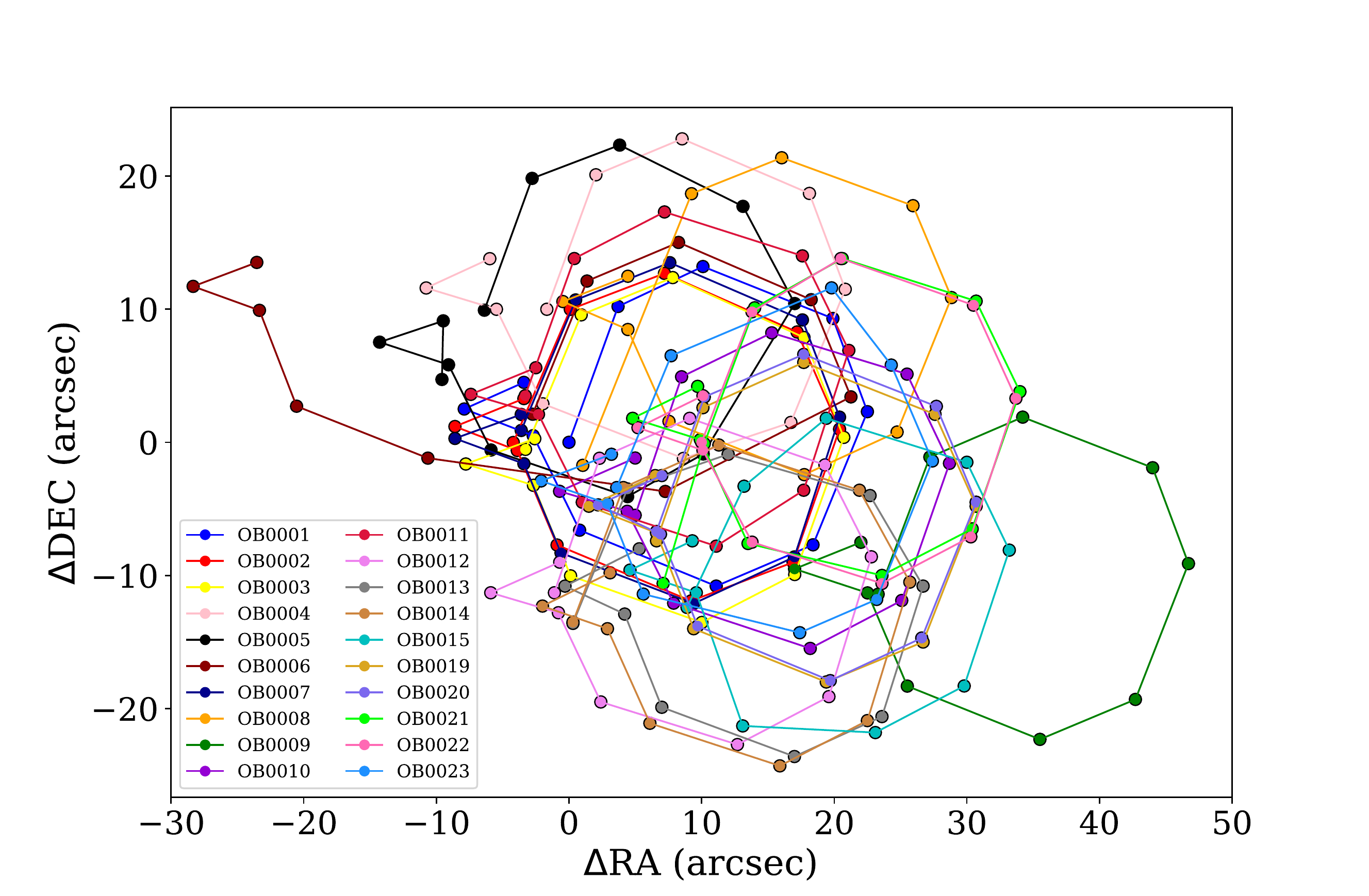}}
  \caption{Dither patterns of the 20 OBs used in this analysis plotted with different colors. The offsets were obtained using two stars as reference and refined using the cross-correlation technique. We took the first image of OB0001 as the reference point (0,0). The error bars are too small (smaller than 0.1~arcsec) to be seen in the figure.}
  \label{dithering_patterns}
\end{figure}

In Fig.~\ref{dithering_patterns} we overplot the dithering patterns for all the OBs finally used in the generation of the final image with different colors, using the pointing of the first image of OB0001 as a common reference.
The program {\sc imcombine}, especially written by one of the authors to combine \mbox{near-IR} images, takes as input a list of the single images to be combined, the initial offset determination between them, a bad-pixel mask (facilitated for us by the PI of CIRCE), and a previously computed dark image. In addition, the program is able to compute an object mask, which is initially set to zero and initialized with the detected sources at the end of the execution of the program. In subsequent executions of {\sc imcombine}, the object mask was employed and conveniently updated. The different manipulations carried out by this program are listed below.
\begin{enumerate}
    \item Normalize individual (image$-$dark) frames, dividing by the median of the full image, excluding the already detected objects (none in the first execution of the program, and those detected in step 6 below in subsequent executions), and avoiding also the pixels that are included in the bad-pixel mask.
    \item Compute superflat from the mean of the stacking of the normalized images, excluding pixels in the object mask.
    \item Compute individual (image$-$dark) / superflat.
    \item Sky subtraction in (image$-$dark) / superflat: the sky level is computed as the median of the signal in each channel of the CIRCE detector, avoiding pixels included in either the bad-pixel or the object masks.
    \item Combine sky-subtracted (image$-$dark) / superflat frames: for each pixel, the median value of all the pixels in the available frames is computed. Two auxiliary images are also generated, storing the median absolute deviation and the number of pixels from the individual frames employed for each pixel in the final image.
    \item Generate the object mask using SExtractor \citep{1996A&AS..117..393B}. We performed some tests to determine the optimal SExtractor detection parameters that maximized the number of detected objects while minimizing the number of false detections.
\end{enumerate}

\begin{figure}[t]
\centering
  \resizebox{\hsize}{!}{\includegraphics{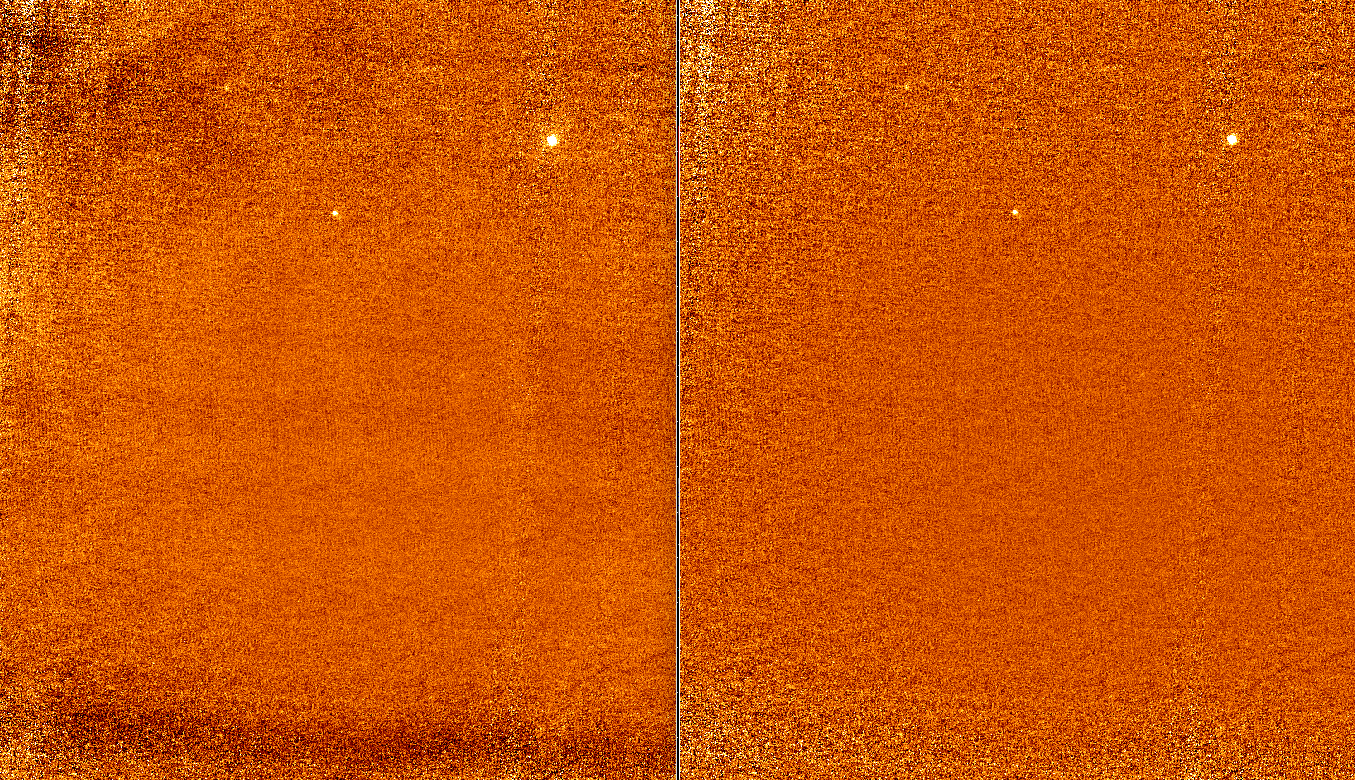}}
  \caption{Comparison between a final OB image before (\textit{left panel}) and after (\textit{right panel}) applying the nebular filter. Although there are still noisy areas around the borders as a consequence of the dithering and the bad cosmetic of the detector, many dark spots disappear. This flattens the final image and allows us to perform reliable photometry.}
  \label{comparison_OB22}
\end{figure}

The careful examination of each combined OB image revealed that the image background was far from perfect. It exhibited some deviations from zero at intermediate scales. These imperfections were more noticeable at the borders of the combined images, indicating that the bad-pixel mask was not too aggressive (i.e., it did not contain too many bad pixels), which in turn could produce systematic variations in the computation of the superflatfield in these image regions. Instead of dramatically increasing the list of bad pixels, which would have led to a significant loss of the effective field of view in the combined images, we decided to remove the background variations making use of the {\sc nebuliser} program that is included in the CASUTools package\footnote{\url{http://casu.ast.cam.ac.uk/surveys-projects/software-release}}. Although initially developed to fit spatially varying nebulosity in images, this software program works very well for modeling varying backgrounds at medium and large scales through the use of a series of iterative sliding median and mean filters that are applied to each axis or to both simultaneously \citep{Irwin}. In Fig.~\ref{comparison_OB22} we compare the result of applying this correction in one of the combined OB images. It is clear that the good tracking of background variations allows removing a significant amount of the dark structures of the image. This effect is even more pronounced when comparing the result of combining the 20 final OB images before and after applying the nebular filter. Hence, after applying {\sc nebuliser,} the image is flatter. This reduces the errors in the photometry in the final NB1257 image.

\subsection{Step 3: Final NB1257 image}

The final step 3 in the data reduction process (see Fig.~\ref{DR_diagram}, right panel) consisted of the combination of the final images of each OB obtained after the execution of step 2, with the goal to build the final NB1257 image.
Although the initial data set consisted of 23 OBs, just a simple visual inspection of the images corresponding to OB0016, OB0017, and OB0018 revealed that they exhibited an unexpectedly high noise level.
We found no reasonable explanation for this, apart from the fact that these three blocks correspond to observations carried out in two consecutive particular nights. It is therefore likely that some problem related to the electronics of the instrument was responsible for it. 
We performed an exhaustive analysis of the noise of the images to select only the best OBs for the final combination. Further details about the method we followed in this case are provided in Appendix~\ref{circe_errores}. Because this analysis revealed what we previously saw by eye, we decided to remove the affected OBs before the final combination (Table~\ref{tableOB}, Col.~8). 
The dithering technique makes the good alignment of the individual images crucial for the seeing measurements. The mean seeing during the observations was \mbox{0.76 $\pm$ 0.14 \arcsec}, and we show in  Col. 6 of Table~\ref{tableOB} the seeing values of the final image of each OB. The observations were spread over a year, and although there were seeing variations, we demonstrate in Appendix \ref{circe_errores} that in the case of our images, the noise dominates the signal-to-noise ratio (\mbox{S/N}) of the final NB1257 image. For this reason, we decided to use  only the OBs that contribute to increasing the S/N of faint sources in the final combination.

After discarding the unusable images, the total exposure time of the remaining 20 OBs amounted to 18.3 hours.
Before piling up the useful OBs, we discovered that some of them exhibited a non-negligible counter-clockwise rotation of 3 degrees. For the final combination, we therefore made use of the IRAF tasks {\sc rotate}, {\sc ccdproc,} and {\sc imcombine}, which allowed the rotation correction, the determination of the relative offsets between individual images, the cropping of some noisy image borders, and the final image combination. The last step was to reverse the image in the X direction.

\begin{figure*}[ht]
  \resizebox{\hsize}{!}{\includegraphics{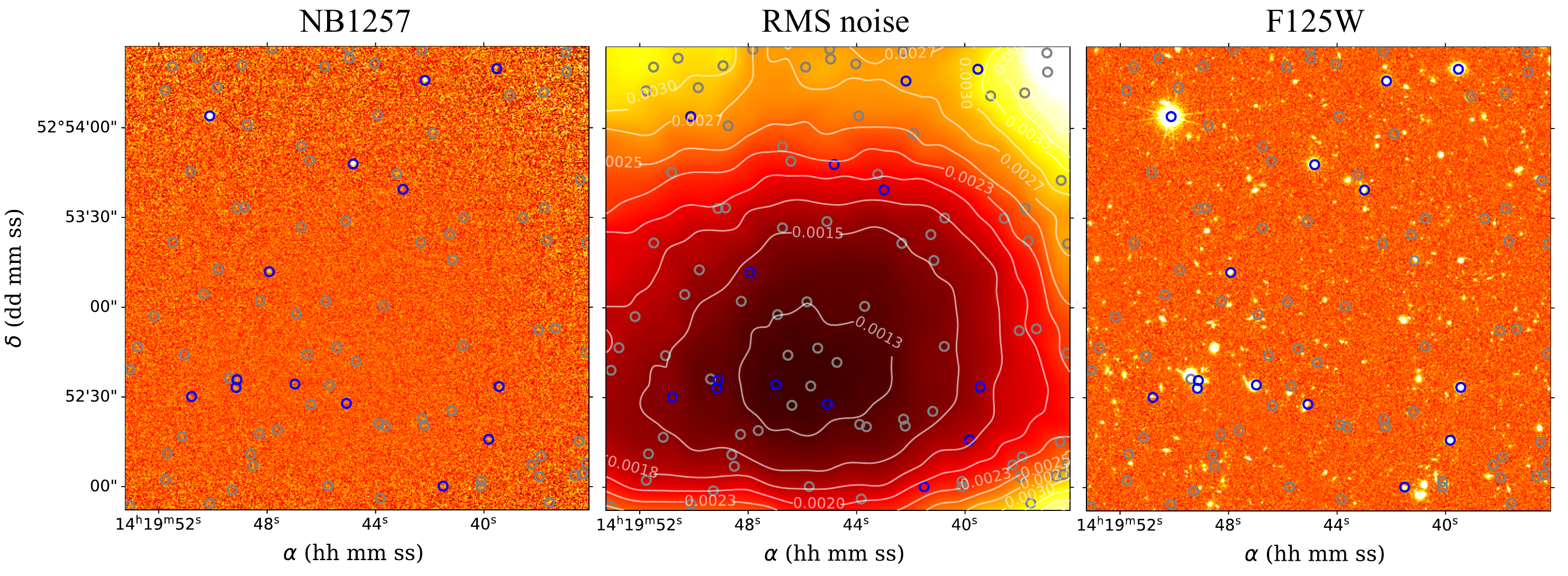}}
  \caption{Field of view within the EGS field. The positions of the detections are plotted with open circles: blue for already identified objects at lower redshifts (\mbox{subsample 1}) and gray for sources without an HST counterpart (\mbox{subsample 2}). \textit{Left panel}: GTC/CIRCE deep image (NB1257 filter) with 18.3~h of exposure time. \textit{Middle panel}: RMS noise contour map of the NB1257 image in ADU/s. \textit{Right panel}: \mbox{HST/WFC3} image (F125W filter) with 25.9~h of exposure time. Orientation: North is up, and east is to the left. Size: $2.6 \times 2.6$~arcmin$^{2}$. }
  \label{comb19}
\end{figure*}

The final deep NB1257 image is shown in Fig.~\ref{comb19} (left panel). 
Our final image exhibits cosmetic defects, especially at the borders. These are regions with low S/Ns that are located at the edge of the FOV, which are caused by the dithering pattern. Nevertheless, an important fraction of the image is useful for our scientific purposes, although it is worth noting that not all parts of the image were exposed equally, and different depths were reached. This effect is clearly visible in the RMS noise map (Fig.~\ref{comb19}, middle panel) where the less noisy region (i.e., the deepest) is found at the center of the image, while the RMS increases significantly at the edges. The RMS noise image was obtained using the \textit{photutils.background} package of {\sc photutils}\footnote{\url{https://photutils.readthedocs.io/en/stable/}}, which provides several tools for detecting and performing photometry of astronomical sources \citep{larry_bradley_2020_4044744}. 
Although the field of view of each CIRCE pointing is 3.4~$\times$~3.4~arcmin$^{2}$, the combined OB images were trimmed before the final combination since the borders of the images are severally affected by the bad cosmetic
of the detector,  and the area covered by the final NB1257 image is 2.58~$\times$~2.58~arcmin$^{2}$.
For comparison, and because we chose it as our BB image, we also show in Fig.~\ref{comb19} (right panel) the image taken in this region by the HST with the \mbox{WFC3/F125W} filter. 

\section{Image characterization and source detection}\label{candidateidentification}

In this section, we describe several steps that are required before the candidate identification. First, we explain the process of PSF matching and flux calibration between the broad and the narrow images. Then, we determine the optimal parameters for the sources extraction, and finally, we study the contamination and detection completeness of the NB1257 image.


\subsection{PSF matching and flux calibration}

Before the source detection was carried out and the photometry was measured, the NB1257 and F125W images must have the same size and pixel-by-pixel coordinates to ensure that we measured fluxes at the same positions and apertures in both images. First, we used the IRAF task {\sc magnify} on the \mbox{WFC3/F125W} HST image to match the pixel step size of the NB1257 image (0.1~$\arcsec$/pixel). Next, we aligned both images using the bright sources of the field as a reference.

We smoothed the \mbox{WFC3/F125W} HST image to the spatial resolution of the NB1257 image (FWHM~$\sim $~0.8~$\arcsec$) using the \textit{photutils.psf} package \citep{larry_bradley_2020_4044744}. In particular, we combined a stack of the brightest but unsaturated isolated stars to build an empirical PSF model for each image. Then, we used the PSFs to generate a matching kernel between both images using the ratio of Fourier transforms. Finally, we carried out a 2D convolution of the HST image and the generated kernel to achieve the PSF matching.

To obtain the zeropoint needed to transform ADU/s of our NB1257 image to AB magnitudes, we took the F125W band as reference, and we assumed that the mean color for the brightest stars was zero, as done in \cite{2008ApJ...677..169V}. The flux calibration of the NB1257 image was achieved using the \mbox{HST/G141} grism spectra (\mbox{3D-HST} survey) of two stars in the field. We computed their synthetic magnitudes using two different Python packages, {\sc synphot}\footnote{\url{https://synphot.readthedocs.io/en/latest/}} and {\sc pyphot}\footnote{\url{https://mfouesneau.github.io/docs/pyphot/}} , and we obtained consistent results with the two independent methods (the relative error between the estimated zeropoints was lower than $\sim0.0003 \%$).


\subsection{Source detection}

After matching and calibrating the images, we carried out the source extraction. We ran the Python code {\sc SEP}\footnote{\url{https://sep.readthedocs.io/en/latest/}} \citep[\textit{Source Extraction and Photometry} in Python: ][]{kyle_barbary_2015_15669}, which is derived from the SExtractor code base \citep{1996A&AS..117..393B}, on the NB1257 image for the object detection (\textit{sep.extract} package). We performed some tests to reach the optimal parameters that push the detections as deep as possible while minimizing the false detections. The pixel-by-pixel detection threshold (\texttt{thresh}) was $1.7~\times$~RMS noise image  (\texttt{err}). The minimum number of pixels required for an object (\texttt{minarea}) was fixed to match the FWHM area of the PSF and thus set to 50~pixels. The filter treatment (\texttt{filter\_type}) was set to \texttt{matched} to ensure a better detection of faint sources in areas of varying noise. The final sample is composed of 97 sources detected in the NB1257 filter, whose coordinates are marked with open circles in Fig.~\ref{comb19}. To check the heterogeneous depth of the final image, we compared our RMS image with the one generated using the background estimation of the \textit{sep.background} package, and no significant difference was found (relative error lower than $\sim$~0.02\%).

In an attempt to search for extremely faint objects, we searched all the detections using relaxed constraints. 
We tested the source extraction using different values of \texttt{minarea}, but values lower than 50~pixels (the smallest area that we can resolve) caused a considerable increase in the number of unreliable detections. For instance, we obtained 157 detected objects when using a value of 40~pixels and 287 objects when \texttt{minarea} was set to 30~pixels. On the other hand, if we varied the \texttt{thresh} parameter to values of 1.5, 1.2, and 1.0, we detected 132, 154, and 246 objects, respectively (with a fixed value of 50~pixels for the \texttt{minarea}). However, this increase in the total number of detected sources was directly reflected in the number of false detections.
Therefore, we considered the previous values as the optimal parameters for the source detection, resulting in a total of 97 detected sources. 


\begin{figure*}[ht]
  \resizebox{\hsize}{!}{\includegraphics{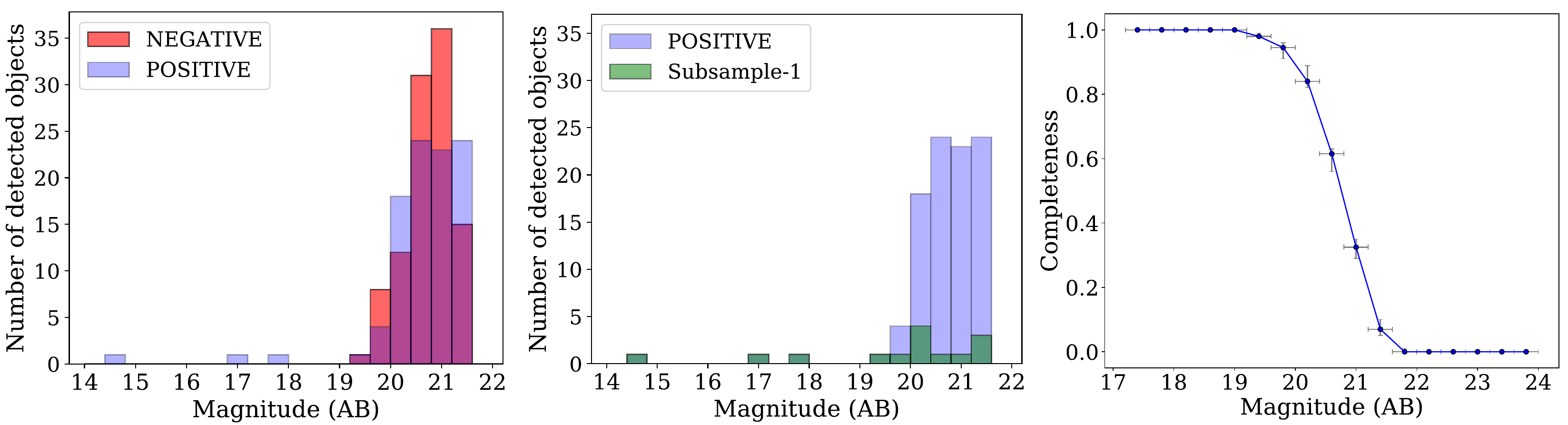}}
  \caption{Contamination and completeness levels. \textit{Left panel}: NB magnitude histogram of objects detected in the negative image (red bars) and in the positive science image (purple bars). \textit{Middle panel}: Histogram with the number of objects detected in the positive science image (purple bars) and the number of them that are already identified objects at lower redshifts (subsample 1, green bars). \textit{Right panel}: Completeness levels as a function of AB magnitude with bins of 0.4~mag. The error bars in the vertical axis represent the 16th and 84th percentiles of the distributions obtained from the simulations.}
  \label{cont_compl}
\end{figure*}

\subsection{Detection completeness and contamination}

In order to obtain the depth and quality image assessment, we computed the contamination and completeness levels. First, we used the segmentation map created from {\sc SEP} to mask the bright sources of the image, and we multiplied the image by $-1$ to create a negative image in which no objects should be detectable.
After running {\sc SEP} on the negative image with the same set of parameters as were used for the detection in the NB1257 image, we compared the number of objects detected in the negative and the science image as a function of the AB magnitude (calculated with the flux of each object extracted from {\sc SEP}). This comparison is shown with histograms in Fig.~\ref{cont_compl} (left panel), where the red bars represent the detections in the negative image and the purple bars the objects detected in the NB1257 image (the positive image). All the objects detected in the negative image are spurious detections. The ratio of these objects and those detected in the positive image in each magnitude bin should represent the contamination levels in the science images. However, the negative image is not fully representative of the NB1257 image because there are magnitude bins where the spurious detections exceed the detections of the science image. We interpreted this as a consequence of the bad cosmetic of the detector. Notwithstanding this fact, we still considered it was useful for determining the magnitudes where we expect more contamination of sources.
The number of objects detected in the negative image with magnitudes fainter than 20.4~mag increases considerably, and therefore, we assumed that this magnitude range presents a higher level of contamination. This implies that a non-negligible number of sources detected in this faint magnitude regime are likely spurious-like detections. Despite the small size of our sample, it is clear that there is no contamination for magnitudes brighter than 19.0~mag. Hence, we conclude that the sources detected in this magnitude range are real detections.

We studied the completeness of the final NB1257 image by injecting artificial stars in simulated RMS images and determining the recovered fraction by running {\sc SEP}. The simulated RMS images were created by measuring the robust standard deviation (see Equation \ref{robustSTD}) around each pixel in the NB1257 image (where previously the bright sources were masked) and generating random Gaussian noise with the same standard deviation. Then, we used the {\sc starlist} and  {\sc mkobjects} tasks of IRAF to generate artificial stars and inject them randomly into each image (100 stars/frame, ranging from 17 to 24 magnitude in bins of 0.4~mag) with a uniform spatial and magnitude distributions.
Finally, the completeness levels shown in Fig.~\ref{cont_compl} (right panel) were derived from the ratio of the numbers of objects detected by {\sc SEP} at the same positions as the simulated stars and the total number of artificial stars. We repeated this recovering process 20 times for each magnitude bin. The errors in the completeness levels are given as the 16th~--~84th percentile interval of the distributions obtained from the simulations, which correspond to $\pm 1\sigma$ for a Gaussian distribution. The completeness level is 100$\%$ at magnitudes lower than 19.0~mag, corresponding to the magnitude range without contamination (see Fig.~\ref{cont_compl}, left panel). Then, it decreases to 50$\%$ at 20.8~mag, and at the 5$\sigma$ limiting magnitude of $\sim$~21.1~mag, the completeness level is 33$\%$. At this stage, we only detected the stars placed in the central region of the image, which corresponds to the deepest region, as shown in the RMS map in Fig.~\ref{comb19} (middle panel). Eventually, the curve reaches zero at 21.6~mag, meaning that beyond this magnitude, we were not able to recover any artificial star in the image. 


\section{Final sample}\label{finalsample}

Aiming to select candidates to Ly$\alpha$-emitters at \mbox{$z = 9.34$}, we built a color-magnitude diagram using the F125W image as our BB image. In addition, we studied our final sample of objects detected by {\sc SEP} describing the criteria applied for the candidate identification.

First, we categorized the 97 sources detected in our NB1257 into two categories depending on whether they present a counterpart in the F125W image. \mbox{Subsample 1} contains the sources with an HST counterpart, and \mbox{subsample 2} includes the sources without a match.
After this classification, \mbox{subsample 1} encompassed 14 objects (four stars and ten galaxies identified in the CANDELS and \mbox{3D-HST} surveys), while \mbox{subsample 2} comprised a total of 83 objects. Within \mbox{subsample 1}, we were able to identify ten galaxies in the redshift range \mbox{$z =$ [0.34 -- 1.53]}. In addition, eight out of the ten galaxies have an available spectroscopic redshift. We report two possible emission-line sources detected in our NB1257: an \mbox{\ion{O}{III$\lambda$4959}} emitter at \mbox{$z_{\rm spec} = 1.53,$} and an \mbox{\ion{O}{III$\lambda$5007}} emitter at \mbox{$z_{\rm phot} = 1.49$} (objects number 50 and 31, respectively). Nevertheless, only spectroscopic follow-up can confirm the latter as a lower redshift contaminant.
In Fig.~\ref{cont_compl} (middle panel) we represent the magnitude histograms of the total sample (purple bars) along with the number of objects belonging to \mbox{subsample 1} (green bars), corresponding to the objects with HST counterparts. These sources represent a small fraction ($\sim$~14\%) of the total sample of 97 objects because most of the objects detected in the NB1257 image belong to \mbox{subsample 2}.

We performed an astrometric calibration of our NB1257 image making use of the software WCS tools\footnote{\url{http://tdc-www.harvard.edu/wcstools/}} and the public coordinates of the objects in \mbox{subsample 1}. We tested this calibration with the 97 detected objects, finding a median deviation of 0.2~$\arcsec$ between our WCS solution and the astrometry of the F125W image. The identification numbers of the whole sample were thus ordered by increasing right ascension.

\subsection{Color-magnitude diagram}

\begin{figure}[t]
  \resizebox{\hsize}{!}{\includegraphics{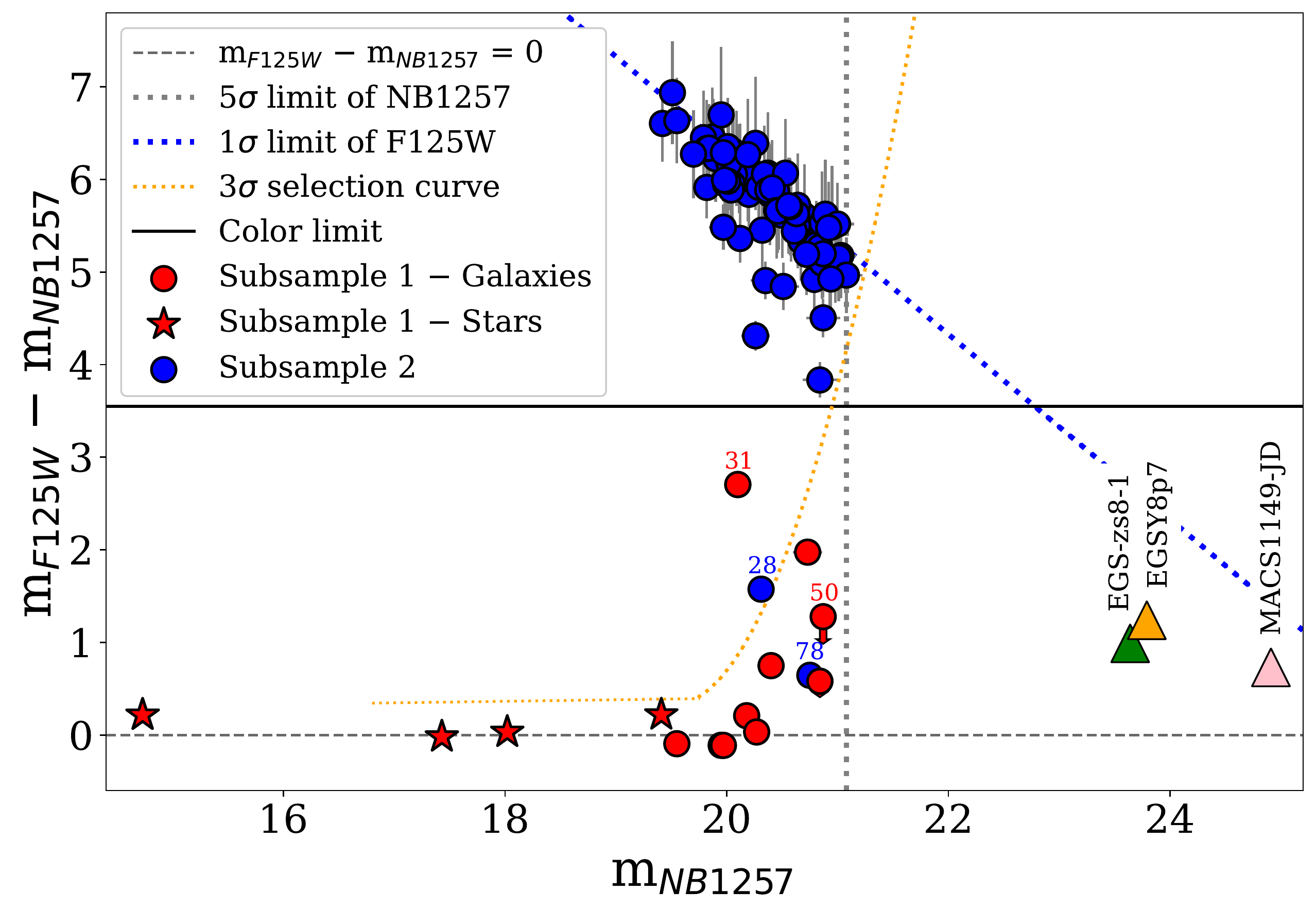}}
  \caption{Color-magnitude diagram of the final sample with magnitudes measured within 0.8$\arcsec$ radius apertures. 
  Red symbols indicate the position of \mbox{subsample 1} (objects with an HST counterpart): filled circles and stars represent the galaxies and stars, respectively. Blue circles represent \mbox{subsample 2} (objects without HST counterpart). The solid horizontal black line shows the upper limit to the color produced by an emission line. 
  The dotted orange line is the selection curve with a level of significance of n$_{\sigma}$~=~3.0 \citep{2007PASP..119...30P}. 
  The 5$\sigma$ (1$\sigma$) limiting magnitude in the NB1257 (F125W) filter is marked with a vertical dotted gray (diagonal blue) line. 
  The position that the very high-$z$ galaxies \mbox{EGSY8p7} \citep{2015ApJ...810L..12Z}, \mbox{EGS-zs8-1} \citep{2015ApJ...804L..30O}, and \mbox{MACS1149-JD} \citep{2018Natur.557..392H} would have in our color-magnitude diagram is plotted with an orange, green, and pink triangle respectively.
  The dashed horizontal gray line marks the m$_{\rm F125W}~-~$m$_{\rm NB1257}=~0$ reference to guide the eye.
  The identification number of some objects (discussed in more detail in the text) is provided. The red arrows indicate the color of the sources in \mbox{subsample 1} using the centroid of the objects in the F125W image instead of the position of the NB1257 detections to measure the BB fluxes. The arrow is discernible only for the faintest source.}
  \label{color_magnitude}
\end{figure}

The key step of the NB technique is to select emission-line candidates by the detection of a flux excess in the NB filter due to their Ly$\alpha$ emission \citep{2007ApJS..172..523M,2010ApJ...723..869O,2010MNRAS.403.1611K,2014ApJ...797...16K}. In a first attempt to select candidates, we represent in Fig.~\ref{color_magnitude} the 97 objects detected in the NB1257 image in a color-magnitude diagram.
We used the NB1257 image to detect all objects and then measured at the same position ($\alpha$ and $\delta$, J2000) in the \mbox{WFC3/F125W} HST image (our BB image). 
We compare the magnitude in the F125W filter with the magnitude in our NB1257 filter, distinguishing between the two subsamples with different colors. The magnitudes were defined with a 0.8$\arcsec$ radius aperture, which corresponds to~$\text{twice the}$~FWHM of the PSF. 
We also checked the color-magnitude diagram with slightly different aperture sizes, but did not recover new candidates.
In this type of diagram, the points around the mean color are distributed asymmetrically. Their dispersion increases with NB magnitude.
This dispersion, usually originated from the intrinsic color of each object, for our sample is mainly due to the large uncertainties in the flux measurements of faint sources.
In our case, the objects with magnitudes in the NB1257 filter fainter than $\sim$~20.0~mag deviate significantly with respect to the mean color. 
It is possible that different sensibilities in the two filters cause the asymmetry in the diagram (the flux in the NB1257 filter might be overestimated). 
Moreover, incorrectly centered apertures in the BB image might generate a systematic effect in the color. We tested this aftermath by measuring the F125W fluxes in the centroid of the sources belonging to \mbox{subsample 1} (i.e., the sources with a counterpart in the F125W band) and computing the m$_{\rm F125W}~-~$m$_{\rm NB1257}$ color again. 
The color change is only relevant for object number 50, which is shown with a red arrow in Fig.~\ref{color_magnitude}.
When we use the position of the NB1257 detections to measure the fluxes in the BB band, we underestimate the F125W fluxes (and thus overestimate the corresponding colors). However, when the offset between the center of the apertures is relatively small (smaller than $\sim$~0.4$\arcsec$), which is the case for most objects, this effect is negligible in the color-magnitude diagram. For the faintest sources alone, the position of the NB1257 detection is significantly different from the centroid of the source measured in the F125W band.
Only object 50 exhibits a relevant discrepancy in color (see Fig.~\ref{color_magnitude}).
We took this outcome into account, but it did not affect the candidate selection in the end.

The errors of the NB1257 (F125W) fluxes were calculated by measuring photometry in the NB1257 (F125W) RMS noise image with the same size aperture. The errors for \mbox{subsample 2} are quite significant, but are almost negligible for \mbox{subsample 1.} The reason is the nondetection in the F125W band. 
The color that a candidate can have due to an emission line is limited by the ratio of the effective widths of the BB and the NB filter \citep{2007PASP..119...30P}. In our color-magnitude diagram, this upper limit (Fig.~\ref{color_magnitude}, solid horizontal black line)
clearly divides our sample into two groups (red and blue dots), corresponding to our two subsamples. 
Moreover, most of the sources in \mbox{subsample 2}  follow the diagonal cloud of the 1$\sigma$ limit of F125W data (dotted blue line). All the sources that are offset from the diagonal trend were identified in the aperture photometry as contaminated by the flux of a nearby object, which would explain their position in the diagram. 
\begin{figure}[t]
  \resizebox{\hsize}{!}{\includegraphics{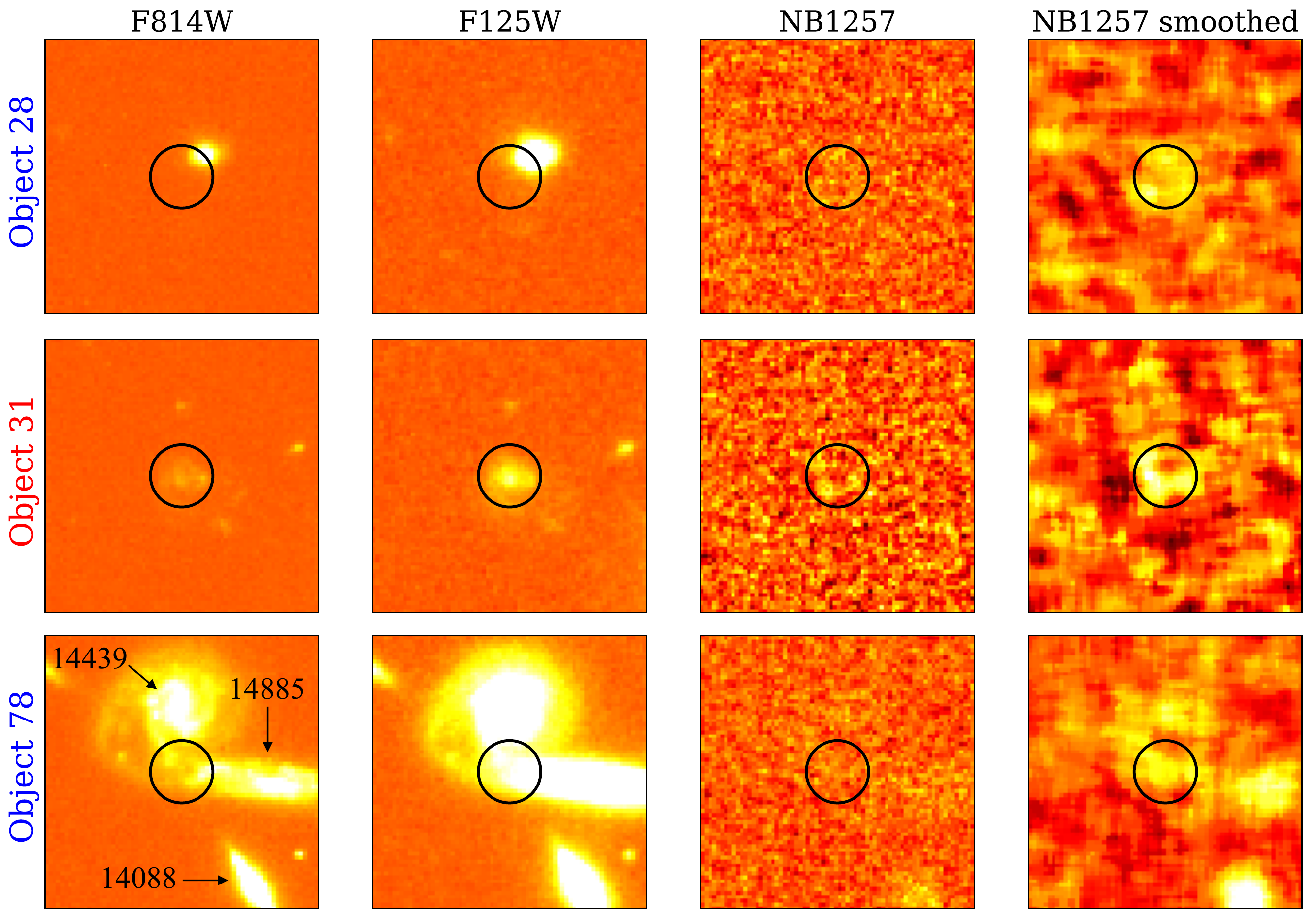}}
  \caption{Sample stamp images of objects 28, 31, and 78 in the HST \mbox{ACS/F814W} and \mbox{WFC3/F125W} bands, our NB1257 filter, and a median filtered version of the latter (NB1257 smoothed, computed only for displayed purposes). The 0.8$\arcsec$ radius aperture is plotted with a black circle at the center of the stamp, corresponding to the positions of the detection in the NB1257 image. The sources nearby object 78 are marked with arrows and the 3D-HST identification number. Orientation: North is up, and east is to the left. Size: $7~\times~7$~arcsec$^{2}$. }
  \label{candidates}
\end{figure}
We expect redder colors as the NB1257 magnitude increases, until the 5$\sigma$ limiting magnitude of $\sim$~21.1~mag is reached (Fig.~\ref{color_magnitude}, vertical dotted gray line). It is worth noting that red colors (i.e., faint BB fluxes with respect to the NB signal) imply large equivalent widths \citep{2008ApJ...677..169V}. 
The equivalent widths were computed following the approach described in \cite{2006A&A...460..681H} \citep[see also][]{2007PASP..119...30P}.
As expected, the estimated EWs of most of the sources belonging to \mbox{subsample 2} (except for objects 28 and 78) resulted in negative values because the color limit was exceeded. The extended sources of \mbox{subsample 1} cover a wide range of observed equivalents widths: \mbox{EW$_{\rm obs}$ $\sim$ [23 -- 797]~$\AA$} (with the exception of object 31).

LAEs at \mbox{$z = 9.3$} should show a flux excess in the NB1257 with respect to the F125W band. The selection curve defined by \cite{2007PASP..119...30P} was employed to identify LAE candidates above the curve (Fig.~\ref{color_magnitude}, dotted orange line). The level of significance was set to n$_{\sigma}~=~3.0$ and the curve was extrapolated using a polynomial fit because the data do not contain enough points to directly calculate the dispersion in all the NB1257 magnitudes intervals. 
We found two sources in the region of LAE candidates: objects 28 ($\alpha$~=~14h~19m~41.15s, $\delta$~=~+52$\degr$~53$\arcmin$~15.9$\arcsec$) and 31 ($\alpha$~=~14h~19m~41.51s, $\delta$~=~+52$\degr$~52$\arcmin$~00.2$\arcsec$). 
Figure~\ref{candidates} shows their stamps in the NB1257 filter and the \mbox{ACS/F814W} and \mbox{WFC3/F125W} HST bands. Moreover, the NB1257 image was smoothed by a median filter for visual purposes because this image highlights the signal better. The position of the NB1257 detection is marked with a black circle (radius of 0.8$\arcsec$).
Object 31 has a counterpart in the F125W band, and the position of the NB1257 detection corresponds to the centroid of the source in the BB image with an angular separation smaller than 0.1$\arcsec$, meaning that its color is not affected by a systematic effect due to incorrectly centered apertures (see Fig.~\ref{color_magnitude}, its red arrow is too small to be noted).
The estimated rest-frame equivalent widths for objects 28 and 31 were EW$_{0}$~$\sim$~45~$\AA$ and 204~$\AA$, respectively. This agrees well with the limit (Ly$\alpha$ EW$_{0}$~$\gtrsim$~20~$\AA$) derived from classical LAEs detected with NB filters \citep{2016ARA&A..54..761S,2020ARA&A..58..617O}. Furthermore, the EW of object 31 is consistent with the results found for the luminous galaxy CR7 \citep{2015ApJ...808..139S,2019MNRAS.482.2422S} at \mbox{$z \sim 6.6$} (EW$_{0}$~$\sim$~210~$\AA$).

\begin{table}[t]\centering \caption[3galaxies]{Properties of the galaxies nearby object 78}
\begin{threeparttable}
\begin{tabular}{l c c c c}
\toprule\midrule
\phantom{X}ID & $\alpha$ (J2000) & $\delta$ (J2000) & z$_{phot}$ & z$_{spec}$\\
\midrule
14439 & 14h 19m 49.40s   & +52$\degr$ 52$\arcmin$ 37.7$\arcsec$  & 0.29 & $-$\\ 
14885 & 14h 19m 49.14s   & +52$\degr$ 52$\arcmin$ 35.8$\arcsec$  & 0.69 & 0.5093\\  
14088 & 14h 19m 49.19s   & +52$\degr$ 52$\arcmin$ 33.1$\arcsec$  & 0.39 & 0.4539\\ 
\bottomrule\addlinespace[1ex]
\end{tabular}
\tablefoot{Identification number (ID) and physical parameters obtained from \mbox{the 3D-HST} catalog. No spectroscopic redshift is available for galaxy ID-14439.}
\end{threeparttable}
\label{table_3galaxies}
\end{table}

Object 78 ($\alpha$~=~14h~19m~49.39s, $\delta$~=~+52$\degr$~52$\arcmin$~36.2$\arcsec$) is one of the two sources of \mbox{subsample 2} that are located below the color limit. It shows a slightly NB1257 excess, but not enough to be placed above the selection curve. 
Although the signal of the object is observed in the NB1257 smoothed stamp, the F814W band reveals that the position of the object does not match the center of any foreground object (see Fig.~\ref{candidates}). This is a particular case where the presence of an extremely nearby object led to a failed source detection. This was the reason why object 78 was initially classified into \mbox{subsample 2}. Its color is extremely affected by the presence of nearby sources due to contamination by flux during the aperture photometry. We provide the coordinates and redshifts (photometric and spectroscopic) of the three sources nearby object 78 
in Table~\ref{table_3galaxies} in order to investigate whether an emission line might affect this measurement. Nonetheless, according to the available redshifts of these galaxies, no emission line lies within the wavelength range covered by our NB1257 filter.

By comparison with other studies, we included the expected location of three very high-$z$ galaxies from the literature in our color-magnitude diagram (Fig.~\ref{color_magnitude}, triangles). 
The F125W and NB1257 magnitudes were computed using the published Ly$\alpha$ fluxes and EWs. In particular, we derived the expected colors of the spectroscopically identified galaxies MACS1149-JD at \mbox{$z = 9.1096$} \citep{2018Natur.557..392H,2018ApJ...854...39H,2020ARA&A..58..617O,2021MNRAS.505.3336L}, EGSY8p7 at \mbox{$z = 8.683$} \citep{2015ApJ...810L..12Z} and EGS-zs8-1 at \mbox{$z = 7.7302$} \citep{2015ApJ...804L..30O}. 
We conclude that all these galaxies, even the two luminous galaxies in the EGS, are beyond the limiting magnitude of this work and therefore would be undetectable in our NB1257 image.


\subsection{Final candidate analysis} \label{candidateconstraints}

We decided to conduct a more quantitative analysis based on complementary parameters to verify the robustness of the detections.
This allowed us to determine not only possible candidates, but also whether the detections are real or spurious. For this purpose, we took the following complementary criteria into account:

\begin{enumerate}

    \item The position in the color-magnitude diagram: LAE candidates should be placed above the selection curve and below the color limit.
    \item The nondetection in the ACS/F606W and ACS/814W bands: We do not expect \mbox{$z \sim 9.3$} LAEs to be detected in any of the filters blueward of the Ly$\alpha$ line redshifted to 1.257~$\mu$m. 
    \item The counterpart in the \mbox{WFC3/F160W} and \mbox{IRAC--3.6~$\mu$m} and 4.5~$\mu$m bands: The ultraviolet (UV) continuum from luminous Lyman-break galaxies (LBG) at very high $z$ could be observed in filters redward of the J band \citep{1999ApJ...518..138H,2017MNRAS.469.2646C}.
    \item The shape of the objects derived from the ellipse parameters extracted from {\sc SEP} (major semi-axis, minor semi-axis, and position angle): Because very high-$z$ LAEs tend to be compact sources \citep{1996Natur.383...45P,2011ApJ...735....5F,2020ARA&A..58..617O}, elongated sources are prone to be spurious LAE identifications.
    \item The shape of the member pixels of each object in the segmentation map extracted from {\sc SEP}: For the same reason as before, we are cautious with extended sources that might be generated by noise structures. 
    \item The position in the RMS noise map: The edges of the NB1257 image are high-noise regions (see Fig.~\ref{comb19}, middle panel). Objects detected near the borders are therefore likely spurious detections, unless they are bright sources.

\end{enumerate}

A careful visual inspection of the 97 objects in all the available bands was performed independently by three of the authors. However, we did not select any candidate by the dropout technique. All the sources of the sample either present a counterpart in the bands blueward of the NB1257 band (criterion 2) or do not show a significant signal in any band, including the H and 3.6~$\mu$m bands in the red part of the spectrum (criterion 3). 

Considering these criteria, we conclude that \mbox{subsample 1} includes the low-$z$ contaminants of the sample, which are all the objects that were previously detected and identified by other surveys as sources at redshifts lower than \mbox{$z \sim 9$}. These sources failed color criterion 2 because all of them are clearly detected in the HST images (see Fig.~\ref{comb19}, blue circles). 
On the other hand, \mbox{subsample 2} contains the objects that are detected in our NB1257 filter, but not in any other band (see Fig.~\ref{comb19}, gray circles). 
We individually studied these objects, and based on the criteria described above, unfortunately, none of them were reliable. 
Because they were not detected in the blue bands, they satisfied criterion 2. The lack of signal in the F160W band could be produced by a faint UV continuum, evading criterion 3. They are distributed throughout the image, and some of them are located in the deepest part.
However, we searched for compact sources that resembled the appearance of other LAE candidates from the literature, but most of these sources were elongated. Their segmentation maps exhibited a scattered rather than concentrated signal. Above all, as shown in Fig.~\ref{color_magnitude}, they did not satisfy the required position in the color-magnitude diagram. 
We hereby state that their NB1257 flux excess was boosted by the noise. These detections were considered spurious-like sources. 
This result is the consequence of pushing the objects extraction to the limit in an exhaustive attempt of detecting faint sources.

Focusing specifically on the candidates that were selected by their position in the color-magnitude diagram (objects 28 and 31), we analyzed their nature in further detail. 
Object 28 was initially classified into \mbox{subsample 2} because it was not detected in the F125W band (Fig.~\ref{candidates}). The F814W and F125W stamps show the offset between the position of the detection and the center of the foreground source. As observed for object 78, the contamination by the flux of an adjacent source explains the position of this object in the color-magnitude diagram.
Object 31 belongs to \mbox{subsample 1} and is placed above the selection curve. It is roughly three magnitudes brighter in the NB1257 filter than the F125W. It is clearly detected in all the available bands, although the nucleus is especially bright in the red bands (F125W and F160W). This object was identified as galaxy CANDELS-ID\#13218. Nevertheless, the photometric redshift available for this source is \mbox{$z = 1.49$}, suggesting that the \mbox{\ion{O}{III$\lambda$5007}}  emission-line might be the cause of its flux excess in the NB1257 filter. In this context, the bright signal observed in the F160W band would be produced by H$\alpha$ emission. Nonetheless, spectroscopic follow-up would be needed to derive firmer conclusions.
Ultimately, the two initial selected candidates were thereby rejected as possible LAEs at \mbox{$z \sim 9.3$}.


\subsection{Complementary candidates from ancillary data} \label{complementary_candidates}

The study of very high-$z$ galaxies is crucial to provide information on the IGM transparency and to identify ionized bubbles.
The EGS field has been intensively studied over the past years, and many LAEs have been detected at \mbox{$z >$ 7} \citep{2015ApJ...810L..12Z,2015ApJ...804L..30O,2016ApJ...823..143R,2017MNRAS.464..469S}, suggesting that the IGM is more transparent in this sky region than in other cosmological fields.

Aiming to give new insight into the IGM transparency in the EGS field, we studied the detectability of the Ly$\alpha$ line by searching for very \mbox{high-$z$} sources in our CIRCE field. To address this point, we used the CANDELS catalog to identify all objects with a redshift solution within the redshift range covered by our NB1257 filter. The CANDELS catalog does not provide the redshift probability distribution P($z$), but different photometric measurements with their corresponding 68\% and 95\% confidence intervals. Within our CIRCE field, we found only two sources with a median photometric redshift measurement (\mbox{$z_{\rm phot\_med}$}) higher than 8, but none of them within the redshift range covered by our NB1257 filter \mbox{($z$ = [9.29--9.38])}. These objects are \mbox{CANDELS-ID\#36056} \mbox{($\alpha$~=~14h~19m~40.23s,} \mbox{$\delta$=~+52$^{\circ}$~52'~06.57'')} with a median photometric redshift of \mbox{$z_{\rm phot\_med} =$ 8.0}, and \mbox{CANDELS-ID\#37659}  \mbox{($\alpha$~=~14h~19m~39.28s,} \mbox{$\delta$=~+52$^{\circ}$~53'~35.20'')} with a value of \mbox{$z_{\rm phot\_med} =$ 9.5}. No spectroscopic redshift is available for these sources. 
Object \mbox{CANDELS-ID\#37659} does not have a counterpart in the \mbox{3D-HST} catalog (the match is not found within a \mbox{1~$\arcsec$ radius)}. However, object \mbox{CANDELS-ID\#36056} corresponds to source \mbox{3D-HST-ID\#17888}, and we found a photometric redshift at a peak probability distribution of \mbox{$z_{\rm peak}$ = 2.0467} when using the \mbox{3D-HST} catalog. Hence, we conclude that \mbox{CANDELS-ID\#36056} is likely an interloper at \mbox{$z \sim$ 2}. The remaining candidate, \mbox{CANDELS-ID\#37659}, was not detected in this work, and the aperture photometry reveals an S/N lower than one in the NB1257 image. 
Because only one candidate is found at \mbox{$z_{\rm phot} >$ 8}, we were unable to stack the data weighted by the P($z$) to investigate the detectability of the Ly$\alpha$ line in the early Universe.
Multiple factors play a role here: the small area covered by our survey, the small wavelength range covered by the NB1257 filter, and the very high redshift of the sources we are searching for. Because of these factors, we have not found complementary candidates at \mbox{$z \sim 9.3$} from ancillary data within
our CIRCE field.


\section{Constraints on the \mbox{$z \sim$ 9} Ly$\alpha$ luminosity function}\label{LFsect}

Many studies have searched for LAEs using NB filters designed to detect sources with a strong emission line, and specifically, targeting the Ly$\alpha$ line \citep{2008ApJS..176..301O,2010A&A...515A..97H,2010ApJ...723..869O,2017MNRAS.469.2646C}. However, typical extremely distant galaxies are usually very faint, which makes the spectroscopic confirmations at very high redshift still scarce \citep{2017ApJ...845L..16H,2018A&A...619A.136M,2021ApJ...914...79T}. 
In this section, we present the depth reached by this work in both flux and luminosity units. We also compute the comoving volume of the survey and study the effect of the cosmic variance for the \mbox{$z \sim 9$} LAE population. Finally, we place an observational upper limit to the Ly$\alpha$ luminosity function at \mbox{$z \sim 9,$} and we derive the number density using the very high redshift Ly$\alpha$ luminosity functions available in the literature thus far.

We first computed the emission-line fluxes of the sample following the procedure specified by \cite{2007PASP..119...30P}.
Assuming all the sources to be at \mbox{$z = 9.34$}, we derived the corresponding Ly$\alpha$ luminosities (L$_{\rm Ly\alpha}$) from the line fluxes of the sources and the corresponding luminosity distance\textsuperscript{\ref{note1}}, which is 9.6~$\times$~10$^{4}$~Mpc. To estimate them, we adopted apertures of radius 0.8$\arcsec$. 
Considering our 5$\sigma$ limiting magnitude, we reached an emission-line flux limit of 2.9$~\times$~10$^{-16}~$erg~s$^{-1}$~cm$^{-2}$, which corresponds to a Ly$\alpha$ luminosity of 3$~\times$~10$^{44}~$erg~s$^{-1}$. Down to this limit, we found no robust LAE candidates consistent with the selection criteria described in Sect.~\ref{candidateconstraints}. Conducting a searching of LAEs at \mbox{$z = 8.8$}, \cite{2014MNRAS.440.2375M} found, with a larger number statistics, luminous candidates with similar Ly$\alpha$ luminosities (L$_{\rm Ly\alpha}$~$\sim$~1.0~$\times$~10$^{44}$~erg~s$^{-1}$) than ours, although none of them was spectroscopically confirmed in the end.

\subsection{Comoving volume}

The total area covered by the NB1257 image is 6.67~arcmin$^{2}$. Assuming the redshift interval \mbox{$z =$ [9.29 -- 9.38]}, defined by the FWHM of our NB1257 filter, this corresponds to a volume\footnote{\label{note1}The comoving volume and the luminosity distance at a given redshift were computed using the \textit{astropy.cosmology} package.} of 1.1~$\times$~10$^{3}$~Mpc$^{3}$. This is an upper limit because the final NB1257 image is not homogeneous in depth (see Fig.~\ref{comb19}, middle panel).
Furthermore, the transmission curves of the filters present a Gaussian shape instead of a box-shaped profile.
This implies that the effective volume of the survey is not a fixed value, but rather depends on the combined continuum plus emission-line flux. 
Thus, a faint source can be detected only if the emission line lies in the center of the filter, whereas the emission from a bright source may be detected even in the wings. 
In these circumstances, bright galaxies can be detected in larger volumes than their fainter counterparts. 
However, because our NB1257 filter is extremely narrow, we do not expect this effect to be especially relevant for our work (a factor $\sim$~1.5 in volume, \mbox{$\Delta$log $\left( \frac{1}{\rm Volume} \right) \sim$~0.2.)}

Surveying an area similar to ours (6.25~arcmin$^{2}$), \cite{2005MNRAS.357.1348W} searched for Ly$\alpha$ emission at \mbox{$z \sim 9$} using the NB technique without success. No LAE at \mbox{$z \sim 9$} was detected by \cite{2008MNRAS.384.1039W} either. They covered roughly twice the total area and reached a flux limit of \mbox{3.7~$\times$~10$^{-18}$~erg~s$^{-1}~$cm$^{-2}$}. 
Even with a much larger area (32\,400~arcmin$^{2}$), \cite{2014MNRAS.440.2375M} proved that identifying these very high-$z$ galaxies is still a challenge. 
%
\subsection{Cosmic variance}

Our analysis might be affected by observational uncertainties due to the cosmic large-scale structure.
To study this effect, we made use of the publicly available IDL code\footnote{\url{https://www.usm.uni-muenchen.de/people/moster/home/download.html}} described in \cite{2011ApJ...731..113M} to compute the cosmic variance as a function of the redshift bin, the area covered by the survey, and the stellar mass (M$_{\star}$) of the galaxies. Assuming that typical LAEs are low-mass (10$^{\rm 8-9}$~M$_{\odot}$) galaxies \citep{2020ARA&A..58..617O}, we found a relative cosmic variance for galaxies with M$_{\star}$~$<$~10$^{\rm 9}$~M$_{\odot}$ of $\sim$~349\%. This high value is consistent with \cite{2011ApJ...731..113M}, who already stated that for high-$z$ observations it is essential to cover a wide field to reduce the effect of the cosmic variance. 
For instance, for the whole EGS (area of 700~arcmin$^{2}$) and for the same redshift interval, the relative cosmic variance for galaxies with M$_{\star}$~$<$~10$^{\rm 9}$~M$_{\odot}$ is $\sim$~94\%. In addition to the small area covered, the geometry of our survey contributes to the high cosmic variance because the ratio of the two observation angles \citep[$\alpha_1$ and $\alpha_2$ in][]{2011ApJ...731..113M} is one, the maximum. This implies that the typical distance between galaxies is small, favoring their correlation. Considering our total area, a ratio of $\sim$~0.3 in the angular geometry would be needed to decrease the relative cosmic variance by a factor of $\sim$~14\%.

\subsection{Luminosity function}

To better understand the evolution of the LAE population, the evolution of the Ly$\alpha$ LF with redshift must be known. Following the nondetections of our survey, we derived an upper limit on the Ly$\alpha$ luminosity function of log(N(>L))~=~$-$3.0~Mpc$^{-3}$ and L$_{\rm Ly\alpha}$~=~3$~\times$~10$^{44}~$erg~s$^{-1}$ at \mbox{$z \sim 9$}. 
Among the previous constraints from the literature, it is
worth noting the faint detection limits reached by \cite{2007A&A...461..911C} and \cite{2008MNRAS.384.1039W} (L$_{\rm Ly\alpha}$~=~1.3$~\times$~10$^{43}~$erg~s$^{-1}$ and L$_{\rm Ly\alpha}$~=~1.0$~\times$~10$^{43}~$erg~s$^{-1}$, respectively) in spite of the small volumes that were surveyed ($\sim$~5~$\times$~10$^{3}$~Mpc$^{3}$ and $\sim$~2~$\times$~10$^{3}$~Mpc$^{3}$, respectively). 
On the other hand, \cite{2014MNRAS.440.2375M} performed a very wide survey (volume~$\sim$~5~$\times$~10$^{6}$~Mpc$^{3}$), which allowed them to place a tight constraint on the bright end of the LF: log(N(>L))~=~$-$6.7~Mpc$^{-3}$ and L$_{\rm Ly\alpha}$~=~6.3$~\times$~10$^{43}~$erg~s$^{-1}$. 
However, they achieved the same depth as \cite{2009MNRAS.398L..68S} (although in this case, the volume covered was $\sim$~1~$\times$~10$^{6}$~Mpc$^{3}$), which was insufficient to detect any robust LAE candidate at very high $z$. 
Our exclusion zone is consistent with the previous observations at \mbox{$z \sim 9$}.
The estimated Ly$\alpha$ luminosities at \mbox{$z = 9.34$} of galaxies \mbox{MACS1149-JD} and \mbox{EGSY8p7} were 4.7$~\times$~10$^{42}~$erg~s$^{-1}$ and 1.9$~\times$~10$^{43}~$erg~s$^{-1}$, respectively. \mbox{EGS-zs8-1} presents the same Ly$\alpha$ flux as \mbox{EGSY8p7}, hence the estimated Ly$\alpha$ luminosity would be the same for both galaxies. As we already mentioned in Sect.~\ref{finalsample}, these galaxies are beyond the depth reached by our work and, in fact, only the surveys conducted by \cite{2007A&A...461..911C} and \cite{2008MNRAS.384.1039W} were deep enough to be able to realistically detect LAEs with the same luminosity as the luminous galaxies of the EGS. No study performed with an NB filter has yet achieved the Ly$\alpha$ luminosity limit needed to detect faint sources such as \mbox{MACS1149-JD}. Nevertheless, not only the depth, but also the volume that is covered contribute considerably to the probabilities of detecting these young galaxies.

In order to study the space density of galaxies as a function of their luminosity, we made use of the Schechter formalism as described in \cite{1976ApJ...203..297S},
\begin{equation}\label{LF}
    \phi (L) dL = \phi^{*} \left( \frac{L}{L^{*}} \right) ^{\alpha} {\rm exp} \left(- \frac{L}{L^{*}} \right) d\left(\frac{L}{L^{*}} \right),
\end{equation}
where $L^{*}$ is the characteristic galaxy luminosity, $\phi^{*}$ is the number density that provides the normalization, and $\alpha$ is the faint-end slope. The current lack of galaxy detections at {\mbox{$z \sim 9$} makes it impossible to derive a Ly$\alpha$ luminosity function at this redshift, which implies that the luminosity distribution of the {\mbox{$z \sim 9$} galaxy population remains unknown. Even at lower redshifts (\mbox{$z \sim 6-7$}), the LFs are mostly obtained by fitting a Schechter function with a fixed $\alpha$ value due to the scarce data of faint sources that hinder the faint-end slope constraint.
The Ly$\alpha$ luminosity function at the highest redshift to date is given by \cite{2014ApJ...797...16K} at \mbox{$z = 7.3$}, with the best-fit Schechter parameters of $L^{*}_{\rm Ly\alpha}$~=~2.7$^{+8.0}_{-1.2}$~$\times$~10$^{42}$~erg~s$^{-1}$ and $\phi^{*}$~=~3.7$^{+17.6}_{-3.3}$~$\times$~10$^{-4}~$Mpc$^{-3}$ for a fixed $\alpha$~=~$-$1.5.
Likewise, we also considered the Ly$\alpha$ luminosity function provided by \cite{2018ApJ...867...46I}, who used the data given by \cite{2014ApJ...797...16K} at \mbox{$z = 7.3$} to calculate the best-fit Schechter parameters with a steep slope of $\alpha$~=~$-$2.5 \citep[in good agreement with ][]{2021ApJ...919..120M}, obtaining values of $L^{*}_{\rm Ly\alpha}$~=~0.55$^{+9.45}_{-0.33}$~$\times$~10$^{43}$~erg~s$^{-1}$ and $\phi^{*}$~=~0.94$^{+12.03}_{-0.93}$~$\times$~10$^{-4}$~Mpc$^{-3}$.
On the other hand, \cite{2007A&A...474..385N} made use of several observed luminosity functions at lower redshift to extrapolate the LF at \mbox{$z = 9.4$}, obtaining the Schechter parameters of log($L^{*}$)~=~42.92~$\pm$~0.24~erg~s$^{-1}$, log($\phi^{*}$)~=~3.96~$\pm$~0.50~Mpc$^{-3}$ for a fixed $\alpha$~=~$-$1.5.
From these LFs, we can compute the number of galaxies per unit of luminosity and volume. 
Using our Ly$\alpha$ luminosity limit, the result states that we would not expect any LAE detection within our comoving volume. If we consider the same volume but deeper observations, for instance, the estimated Ly$\alpha$ luminosities of \mbox{MACS1149-JD} and \mbox{EGSY8p7}, the expected number of sources would be 0.02 and 0.00002, respectively, when using the LF provided by \cite{2014ApJ...797...16K}. 
It should be noted that the result is an approximation because this LF is derived from observations at \mbox{$z = 7.3$} instead of \mbox{$z = 9.3$}, which means that we would be assuming the case of no Ly$\alpha$ LF evolution from \mbox{$z = 7.3$} to 9.3. Notwithstanding, even though the difference in cosmic time between these redshifts is not significantly large (similar to a sample of \mbox{$z = [0.95 - 1.00]$}), the timescale corresponds to the epoch when the formation of the first galaxies and the reionization took place, which makes observations at these redshifts crucial for understanding how the Universe evolved. We expect a significant evolution of the Universe during the first stages, especially of the IGM, because the double-reionization scenario of the AMIGA model predicts a first peak of ionization around \mbox{$z \sim 10$} \citep{2015ApJS..216...13M,2017ApJ...834...49S}. The single-reionization approach, which claims that reionization was a progressive process, also supports this substantial change in ionized hydrogen fraction of the Universe between these redshifts.  
Using the LF extrapolation at \mbox{$z = 9.4$} given by \cite{2007A&A...474..385N} with the same depths as \mbox{MACS1149-JD} and \mbox{EGSY8p7}, the number density increases slightly, obtaining values of 0.06 and 0.002, respectively. These results indicate that both the ongoing and the next surveys need to cover larger fields and reach fainter magnitudes to be able to detect very high-$z$ sources.

The low number of confirmed galaxies at \mbox{$z > 7$} suggests that luminous sources at very high $z$ may be scarce and therefore difficult to detect. 
Recently, \cite{2021arXiv211014474G} confirmed a decrease in bright LAEs at \mbox{$z > 7$} by performing an NB imaging survey with no LAE detections down to L$_{\rm Ly\alpha}$~=~1.6$~\times$~10$^{43}~$erg~s$^{-1}$ in an effective volume of 2~$\times$~10$^{6}$~Mpc$^{3}$.
In addition, the nature of the galaxies causing the reionization remains debated due to the uncertainty of the escape fraction of very high-$z$ sources \citep{2016ApJ...828...71D,2018ApJ...863L...3C,2021MNRAS.503.4242R}.
Especially around \mbox{z $\sim$ 9}, it is unclear if the main source of ionizing photons comes from bright galaxies \citep{2015ApJ...802L..19R} or faint galaxies \citep{2019ApJ...879...36F}. Hence, a larger number of confirmed detections at \mbox{$z \sim 9$} is essential for building a reliable Ly$\alpha$ luminosity function for this redshift.


\section{Summary and conclusions}\label{summary}

This pilot project aimed to study the feasibility of finding Ly$\alpha$ emitters at very high $z$ using the NB filter designed by the ALBA team. For this purpose, we searched for LAEs in a small region within the EGS field with the \mbox{near-IR} camera CIRCE at the GTC telescope and using a NB filter centered at 1.257~$\mu$m. The NB1257 filter was specifically designed and built for this project to target LAEs at \mbox{$z \sim 9.3$} exploiting one of the OH sky lines windows. We performed a careful data reduction process to ensure the best alignment of the frames and to minimize instrumental effects. In addition, we conducted an analysis of the noise level of each individual image to select only the OBs with the best quality for the final combination. 
Finally, with a total exposure time of 18.3~hours, we reached a limiting AB magnitude of $\sim$~21.1~mag in our NB1257 image. 
A total of 97 objects were detected when the source extraction was made. These sources were placed in a color-magnitude diagram, using the F125W band of HST as our BB image, to identify those with a flux excess in the NB1257 filter. 
Two objects were selected above our selection curve as initial LAE candidates at \mbox{$z \sim 9$}. Then, we described several complementary criteria to confirm the robustness of these initial candidates and to analyze all the detections in more detail. 
At last, no candidates were found down to a Ly$\alpha$ luminosity of 3$~\times$~10$^{44}~$erg~s$^{-1}$ by probing a comoving volume of 1.1~$\times$~10$^{3}$~Mpc$^{3}$. Nevertheless, we placed an observational constraint of the Ly$\alpha$ luminosity function at \mbox{$z \sim 9$} by deriving upper limits from the absence of detections. The exclusion zone reported in this work is consistent with previous observations at this redshift and with the luminosity functions available so far.
From these results, we conclude that the combination of both wider and deeper surveys is essential to detect a first sample of \mbox{$z \sim 9$} LAEs. Notwithstanding, the area covered plays an important role not only in the probability of detection, but also because a large field is required to minimize the effects of the cosmic variance in this population of galaxies. 
Our short-term purpose is to reuse the NB1257 filter in a future \mbox{near-IR} camera, which will allow us to continue searching for very high-$z$ galaxies applying the method we developed and described in this paper. The main motivation is to detect a sample of LAE candidates for subsequent follow-up spectroscopy. If their \mbox{$z \sim 9$} nature is confirmed, these sources will be used to study the feasibility of the \mbox{double-reionization} scenario predicted by the AMIGA model.

In this observational test, we finally did not reach the reionization epoch, but nonetheless, in the near future, several surveys will continue studying the early Universe. The arrival of a new generation of telescopes, such as the SKA \citep{2013ExA....36..235M,2020MNRAS.494..600M}, the JWST \citep{2017jwst.prop.1345F}, or the ELT \citep{2020SPIE11447E..25S}, will provide an unprecedented insight into our knowledge about the first galaxies and Population~III galaxies.
MOSAIC will be a multi-object spectrograph for the future 39~meters ELT, which will be the largest optical and \mbox{near-IR} ground-based telescope.
As shown in \cite{2014A&A...569A..78V}, owing to the high sensitivity and good spatial resolution of MOSAIC, this instrument will improve the magnitude limits in the continuum reached by the \mbox{near-IR} spectroscopy of the JWST \citep[see also the simulations performed for the ELT by ][]{2014SPIE.9147E..91D}. Hence, MOSAIC will become an essential instrument in the next years, enabling us to derive the properties of faint galaxies detected by the JWST \citep{2021Msngr.182...33H} and tracing the reionization.
Many questions currently remain unclear, but probing the cosmic dawn is one of the main science cases of these upcoming facilities.


\begin{acknowledgements}
We thank the anonymous referee for the useful feedback that served to improve the original manuscript.
This research was supported by the Spanish Ministry of Science and Innovation (MCIN) under the grants AYA2016-75808-R and RTI2018-096188-B-I00. 
C.C. acknowledges the Universidad Complutense de Madrid and the MCIN for the grant PRE2019-087503 \textit{"Personal Investigador Predoctoral en Formación"}.
The authors thank Stephen Eikenberry, PI of the CIRCE camera, for his help during the preparation of the observations. This work is based on observations made with the Gran Telescopio Canarias (GTC), installed in the Spanish Observatorio del Roque de los Muchachos of the Instituto de Astrofísica de Canarias, on the island of La Palma. 
The authors also thank the support given by Dr. Antonio Cabrera and the GTC Operations Group staff during the observations at the GTC. 
This study makes use of data from AEGIS, a multiwavelength sky survey conducted with the Chandra, GALEX, Hubble, Keck, CFHT, MMT, Subaru, Palomar, Spitzer, VLA, and other telescopes and supported in part by the NSF, NASA, and the STFC. This work is based on observations taken by the \mbox{3D-HST} Treasury Program (GO 12177 and 12328) with the NASA/ESA HST, which is operated by the Association of Universities for Research in Astronomy, Inc., under NASA contract NAS5-26555. This work has been possible thanks to the extensive use of IPython and Jupyter notebooks \citep{4160251}. This research made use of Astropy,\footnote{\url{http://www.astropy.org}} a community-developed core Python package for Astronomy \citep{2013A&A...558A..33A, 2018AJ....156..123A}.
\end{acknowledgements}



\bibliographystyle{aa}

\begin{appendix}


\section{OB noise analysis}\label{circe_errores}

By visual inspection, we realized that some combined OB images looked noisier than others. In particular, the images taken on the night of 03 June 2017 showed a strange pattern in the background, possibly due to interference with other electronic devices. We searched for possible temperature variations of the detector that might explain this behavior, but we did not detect any significant variation. However, we further analyzed the quality of all the combined OB images in order to quantify the level of noise and to ensure that only the OBs that increased the \mbox{S/N} of the detection were ultimately employed in the final combination.

As explained in Sect.~\ref{data}, the combined OB images are the result of combining 11 single images, except for OB0005, OB0021, and OB0023, which were built by combining 12, 12, and 9 exposures, respectively (see Table~\ref{tableOB}, Col.~2). As a consequence of the dithering technique, each combined OB image presents areas with different depths. 
The robust standard deviation (hereafter referred to as RobustStd) gives us an idea of the noise in a region. It is defined by the following expression \citep{10.5555/2578955}:
\begin{equation}\label{robustSTD}
    {\rm RobustStd} = 0.7413 \, \cdot (p75 - p25),
\end{equation}
where $p25$ and $p75$ are the 25th and 75th percentiles, respectively. The use of this robust estimator facilitated the analysis described in this appendix, avoiding the sensitivity of the traditional standard deviation to outliers. This is important considering the bad cosmetics of the CIRCE detector. We calculated the RobustStd of the distribution of pixels in the different areas of each combined OB image. The RobustStd versus the number of individual exposures (Nexp) employed in the combination is plotted in Fig.~\ref{rms_vs_npix} with different colors for each OB. As expected, the RobustStd is higher in the image regions that are covered by single exposures, while its value decreases with increasing number of individual overlapping exposures. This behavior is shared by all the OBs, although the absolute value of the RobustStd at a fixed number of exposures changes.

In order to select only the OBs with a high enough quality for the final combination, we sorted out the 23 OBs from lower to higher RobustStd, and combined them by adding one at a time. In the end, we generated a total of 23 combined images. We computed the \mbox{S/N} of a subsample of objects in all these combined images to determine the optimum number of OBs, which we included in the definitive combination. For this purpose, we carried out aperture photometry to extract the fluxes of the selected sources and used the error images generated by the program {\sc imcombine} to measure the associated noise (see Sect.~\ref{imcombine_section}, item 5, for further details). To generate the error image of each possible combination, we added the errors associated with the employed OB images in quadrature. The subsample of objects selected for this analysis was defined taking their location in the image and a magnitude in the NB1257 filter lower than 20.2~mag into account to avoid sources that are dominated by background noise.
Figure~\ref{SNR_final_plot} shows the variation in \mbox{S/N} when increasingly noisier OBs were added. We started with a single image, the image with less noise (OB0013), and successively added the remaining OB images, sorted by increasing noise. The \mbox{S/N} always improved with more OBs in the combination, except when we included the last 3 OBs. These 3 OBs (OB0016, OB0017, and OB0018) have the highest RobustStd and are therefore noisiest. The \mbox{S/N} clearly decreases away from the maximum, that is, after combination number 20 (marked with a dashed vertical black line). It is worth noting that the brightest star was also included in the subsample and was selected for the analysis, but it is not plotted in Fig.~\ref{SNR_final_plot} because its \mbox{S/N} is far higher than the S/N of the other sources (reaching values of S/N~$\sim$~70), and it never decreased even when the nosiest OBs were included. This implies that this object is so bright that all the other sources of noise are negligible. However, this case is an exception because the final NB1257 image contains no comparable bright source.

Because the inclusion of OB0016, OB0017, and OB0018 did not contribute to an increase in the \mbox{S/N} of the faint sources, we discarded these OBs and decided to combine only the OB images with a value of the log(RobustStd) lower than 0.75. Hence, a total of 20 OBs were used in the final combination that generated the NB1257 image, with an equivalent total exposure time of 18.3 hours.

\begin{figure}[ht]
  \resizebox{\hsize}{!}{\includegraphics{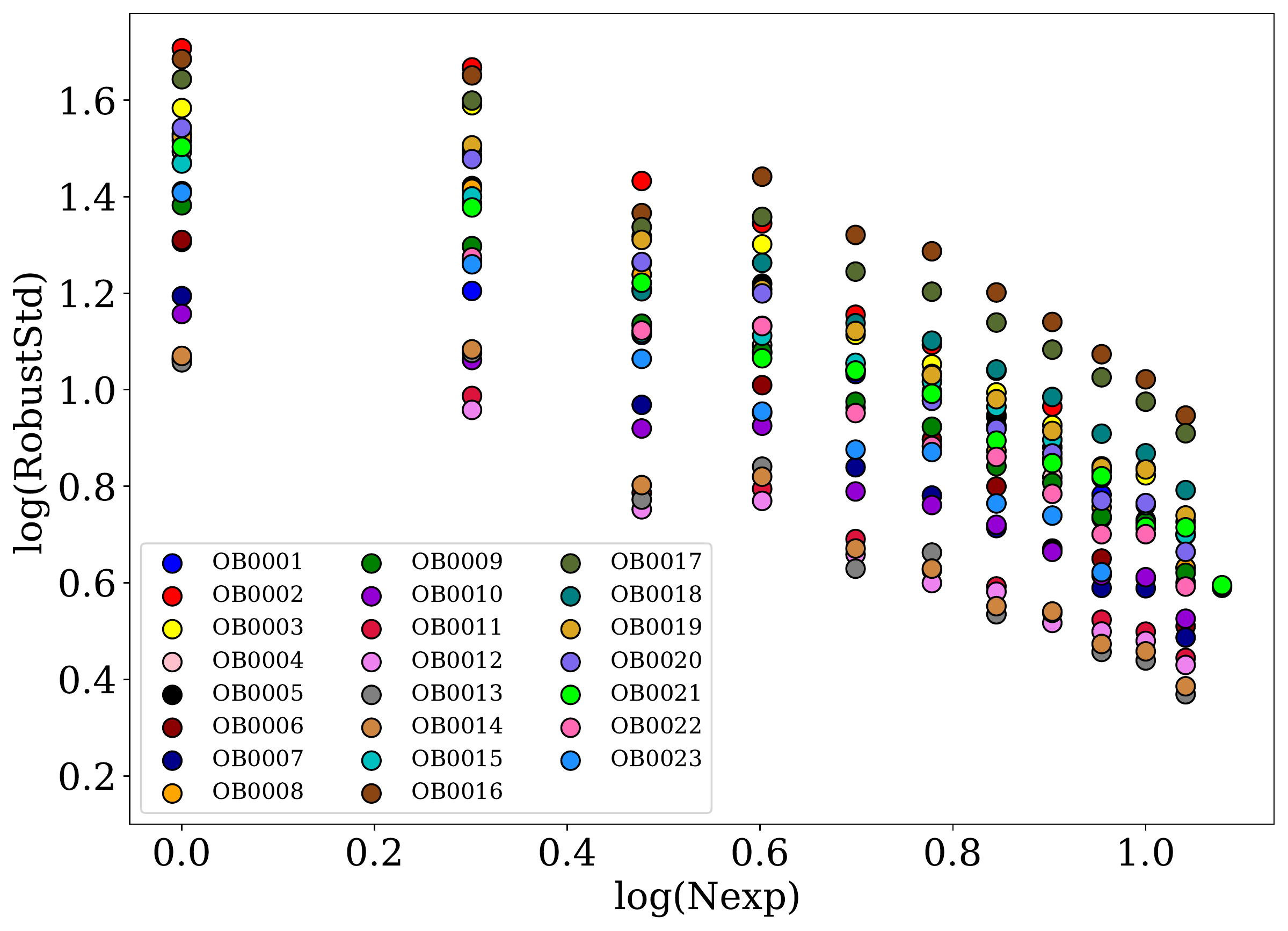}}
  \caption{Robust standard deviation vs the number of exposures employed in the combination for each OB.} 
  \label{rms_vs_npix}
\end{figure}

\begin{figure}[ht]
  \resizebox{\hsize}{!}{\includegraphics{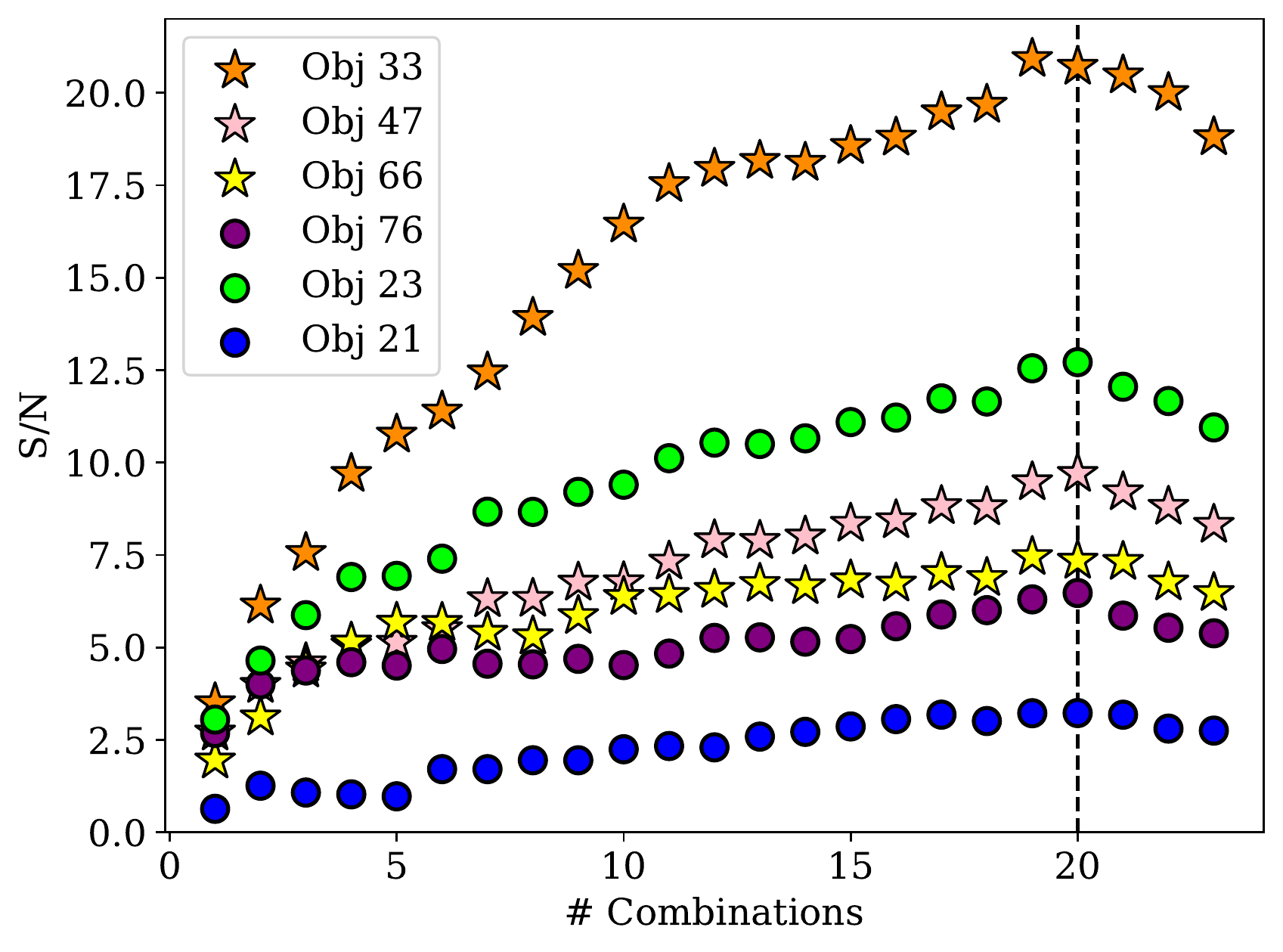}}
  \caption{Variation in S/N according to the number of OBs included in the combination. Filled circles and stars represent the relatively bright galaxies and stars we analyzed, respectively. Different colors stand for different objects. The dashed vertical black line indicates the optimum number of OBs that was used in the final combination. }
  \label{SNR_final_plot}
\end{figure}

\end{appendix}

\end{document}